\documentclass{amsart}
\usepackage{fullpage,amssymb}
\usepackage{epsfig}
\usepackage{url}
\usepackage{color}
\usepackage{pictex}

\newcommand{\dbar}{{\overline{\partial}}}
\urldef{\mcl}\url{mcl@math.arizona.edu}
\def\diag{\mathop{\mathrm{diag}}\nolimits}

\newtheorem{lemma}{Lemma}[section]

\newtheorem{theorem}{Theorem}[section]

\newtheorem{rhp}{Riemann-Hilbert Problem}[section]
\newtheorem{problem}{Existence Problem}[section]

\title[random matrices with external source and algebraic curves]
{Asymptotic analysis of random matrices with external source and a
  family of algebraic curves} 
\author{K. T.-R. McLaughlin}
\address{K. T.-R. McLaughlin:  Department of Mathematics\\ University
of Arizona \\ 617 N Santa Rita Ave, Tucson, AZ 85721
Email address: \mcl }
\date{\today}
\begin{document}
\begin{abstract}
We present a set of conditions which, if satisfied, provide for a
complete asymptotic analysis of random matrices with source term
containing two distinct eigenvalues.  These conditions are shown to be
equivalent to the existence of a particular algebraic curve.  For the
case of a quartic external field, the curve in question is proven to
exist, yielding precise asymptotic information about the limiting mean
density of eigenvalues, as well as bulk and edge universality.
\end{abstract}
\maketitle
\section{Introduction}
\subsection{Random Matrices with Source}

Consider the probability measure 
\begin{eqnarray}
  \label{eq:27}
  \mu_{n}(dM) = \frac{1}{Z_{n}}e^{-n \mbox{ Tr}(V(M) - AM)} dM,
\end{eqnarray}
defined on $n\times n$ Hermitian matrices, $M$.  Here $dM$ denotes
Lebesgue measure on the matrix entries, $dM = \prod_{j=1}^{n} d M_{jj}
\prod_{1 \le j < k \le n}  d \mbox{Re}(M_{jk})d \mbox{Im}(M_{jk})$,
the matrix $A$ is a fixed $n \times n$ matrix, and the parameter $Z_{n}$
is a normalization constant chosen so that (\ref{eq:27}) is a
probability measure.

This defines an ensemble
of random matrices, in which the matrix $A$ plays the role of an
external source.  The function $V$ should be real and grow
sufficiently rapidly that the above measure possesses all finite
moments; it is typically assumed to be a polynomial.

The family of measures described by (\ref{eq:27}) are not invariant
under unitary transformations, and the analysis of eigenvalue
statistics under these measures is in its infancy in comparison to the
so-called ``unitary ensembles'' for which a great deal is known.

Following the work of Br\'{e}zin-Hikami \cite{BH1}, \cite{BH2}, and
P. Zinn-Justin \cite{Zinn-Justin} \cite{Zinn-JustinB}, a team of researchers (
Bleher and Kuijlaars 
\cite{BK1}, \cite{BK2},\cite{BK4} and Aptekarev, Bleher and Kuijlaars
\cite{ABK}) have considered the large $n$ behavior of eigenvalue
statistics under (\ref{eq:27}) from the point of 
view of Riemann--Hilbert problems.

In \cite{Zinn-Justin}, Zinn-Justin showed that the eigenvalues of such
random matrices are in fact a determinantal point process,  and
in \cite{BK1}, the authors showed that this representation may
be described  in terms of the solution of a matrix Riemann--Hilbert
problem, the size of which depends on the number of distinct
eigenvalues of the matrix $A$.

In the present paper we shall assume that the matrix $A$ possesses
two eigenvalues, $-a$ and $a$, with multiplicities $n_{1}$ and
$n_{2}$, respectively (with $n_{1} + n_{2} = n$).  In this case, the
Riemann--Hilbert problem is as follows:
\begin{rhp}
\label{rhp:04}
\begin{enumerate}
\item[(a)] $A$ is analytic on $\mathbb C \setminus {\mathbb R}$. 
\item[(b)] The boundary values of $A$ satisfy 
\begin{equation} \label{RHQHP1}
    A_+(x) = A_-(x)
    \begin{pmatrix}
    1& w_{1}& w_{2}\\
    0& 1&0 \\
    0&0& 1
    \end{pmatrix}, \ \ \mbox{ for } \ z \in {\mathbb R}
\end{equation}
where 
\begin{eqnarray}
w_{1} = e^{- n V_{1}} =  e^{-n \left( V(x) + ax \right)}, \ \ \ w_{2}
= e^{-n V_{2}} = e^{-n \left( V(x) - a x\right)}.
\end{eqnarray}
\item[(c)] As $z \to \infty$, we have
\begin{equation} \label{RHQHPA}
    A(z) = \left(I + O\left(\frac{1}{z}\right)\right)
    \begin{pmatrix}
    z^{n}&0& 0\\
    0& z^{-n_{1}}&0 \\
    0&0& z^{-n_{2}}
    \end{pmatrix}.
\end{equation}
\end{enumerate}

\end{rhp}

In terms of this Riemann--Hilbert problem, the probability measure on
eigenvalues induced by (\ref{eq:27}) may be re-written as follows:

\begin{eqnarray}
\label{probmeas}
d \mu_{n}(x_{1}, \ldots, x_{n}) &=&
\mbox{det } \left(
  K_{n}(x_{i},x_{j}) \right)_{1 \le i,j \le n} \ \ d^{n} x,
\end{eqnarray}
\cite{Zinn-Justin}, \cite{BK1} where now 
\begin{eqnarray*}
K_{n}(x,y) &=&  \left.
\frac{e^{- n ( V(x) + V(y))}}{2 \pi i(x-y)}  \right\{e^{ n a y}\left[
  Y(y)^{-1} Y(x) \right]_{21} \\
& & \hspace{1.2in}
 + e^{- n a y} \left[Y(y)^{-1}
  Y(x)\right]_{31} \Big\}
\end{eqnarray*}

The formula (\ref{probmeas}) is not just a concise representation of the
probability measure, it turns out that all statistical properties of
eigenvalues can be related to $K_{n}$.  For example, 
\begin{eqnarray*}
\mbox{Prob } \left\{  \mbox{no eigenvalues in }
  \left( a, b \right)  \right\} &=& \mbox{ det } \left( 1 -
  \mathcal{K}_{n} \right)_{L^{2}[ (a, b)]}
\end{eqnarray*}
where $\mathcal{K}_{n}$ is the integral operator with kernel
$K_{n}(x,y)$:
\begin{eqnarray*}
\mathcal{K}_{n} f &=&  \int_{a}^{b} K_{n}(x,y) f(y) dy.
\end{eqnarray*}
The representation of the correlation functions in terms of the kernel
$K_{n}$ is due to Zinn-Justin \cite{Zinn-Justin}, and the
representation of this kernel in terms of a Riemann--Hilbert problem
is due to Bleher and Kuijlaars \cite{BK1}.

{\bf Remark:} In the case that the matrix $A$ possesses $p$ distinct
eigenvalues with multiplicities $n_{1}, n_{2}, \ldots, n_{p}$, the
above representations generalize directly, but the associated
Riemann--Hilbert problem is $p+1 \times p+1$.

In \cite{BH1} \cite{BH2}, the authors considered the case that $V$ is
a quadratic.  In the series of papers \cite{BK2}, \cite{ABK}, and
\cite{BK4}, the authors also considered the Gaussian case, but from
the point of view of Riemann--Hilbert problems.  The goal of these
works was to study the behavior of the eigenvalue statistics in the
large $n$ limit.  In
\cite{Zinn-JustinB}, Zinn-Justin also studied the large $n$ limit,
for the case of general $V$.  In that work, one
important issue was the description of the limiting density of states
in terms of a function for which an existence theorem is lacking.
He explains very carefully the analyticity properties and branch-cut
structure required of this function, in the so-called ``one-cut''
case, under the assumption that the density of the eigenvalues of $A$
has a smooth limit when $n$ tends to $\infty$.  

Here we consider the case that $A$ possesses two eigenvalues,
of multiplicities $n_{1}$ and $n_{2}$, and our interest is in the
behavior when $n, n_{1}, n_{2} \to \infty$ so that $\frac{n_{j}}{n}
\to x_{j}$.  The main goal of our work is to provide a new and
explicit characterization of the limiting density of states in terms
of an algebraic curve.  The following is a summary of the results in this paper:
\begin{enumerate}
\item In Section \ref{IdealSitu}, we present a list of conditions (referred to as
  an ``ideal 
  situation'') which, if true, yield a transformation of the
  Riemann--Hilbert problem (\ref{rhp:04}) to a ``normal form'', that
  is, a Riemann--Hilbert problem from which subsequent asymptotic 
  analysis (for $n \to \infty$) is straightforward.
\item In Section \ref{AlgCurve}, we show that if these conditions are
  satisfied, then there 
  exists an algebraic curve whose roots yield the desired
  transformation.  The curve is always of the form 
\begin{eqnarray}
\label{eq:introcubic}
w^{3} - V'(z) w^{2} + \mathcal{C}_{1}(z) w - \mathcal{C}_{0}(z)  = 0,
\end{eqnarray}
where $\mathcal{C}_{1}$ and $\mathcal{C}_{2}$ are analytic functions
of $z$.  
\item For arbitrary polynomial external fields, the algebraic curve
may be determined up to a finite number of free parameters (see
Section \ref{PolyField}).
\item In Section \ref{ExistQuad}, we consider the Gaussian and Quartic cases.
  For the Gaussian case, this algebraic curve has no free 
  parameters, and in the case $n_{1} = n_{2} = n/2$, is equivalent to
  the algebraic curve used by Bleher and Kuijlaars \cite{BK2,BK4} and
  by Aptekarev, Bleher and Kuijlaars \cite{ABK}.
\item For the quartic case, $V = x^4/4$, with $n_{1} = n_{2} = n/2$,
  we prove that for $a$ sufficiently large there exists a choice of
  the free parameters so that the algebraic curve yields the desired
  transformation.  Each of the two relevant roots possesses a single
  branch cut, which is the ``one-cut''case.
\item In Section \ref{sect:rhpanalysis}, in the ``one-cut'' case, we show
  that {\it if the desired 
    transformation exists} bringing the Riemann--Hilbert problem into
  a normal form for subsequent Riemann--Hilbert analysis, then the
  analysis of Bleher and Kuijlaars \cite{BK2} can be applied
  directly.  
\item It follows that if the desired transformation exists,
  then (i) the limiting mean density of eigenvalues exists and is
  described in terms of the algebraic curve, (ii) bulk universality
  holds true, and (iii) edge universality hold true (see Section \ref{results}).
\item The combined results of Sections \ref{ExistQuad} and
  \ref{sect:rhpanalysis} show 
  that for the quartic case $V(x) = x^4/4$, with $n$ even and
  $n_{1}=n_{2}=n/2$, bulk and edge universality still hold true, for
  all $a$ sufficiently large.
\end{enumerate}

{\bf Remark:}  In the analysis of singular limits of $2 \times 2$ Riemann--Hilbert
problems associated to integrable systems and random matrix theory,
there has often arisen an existence problem; a transformation is required satisfying a number
of nonlocal properties, in order to transform to a new
Riemann--Hilbert problem from which subsequent asymptotic analysis is
straightforward (see, for example, \cite{op1}).  The existence of this transformation has been shown
to be equivalent to the existence of an equation of the form 
\begin{eqnarray}
\nonumber
F^{2} + \mathcal{W}(z) F + \mathcal{U}(z) = 0,
\end{eqnarray}
in which (i) the functions $\mathcal{W}$ and $\mathcal{U}$ are
analytic and defined implicitly in terms of the desired
transformation, (ii) all branch points lie on the contour of the
Riemann--Hilbert problem, and (iii) the roots $F_{\pm}(z)$ of this
equation satisfy a variety of inequalities and relations on various
subsets of the contour.

The proofs of existence of these transformations (or equations) are
nontrivial, and require techniques ranging from WKB analysis of odes,
through the theory of level trajectories of quadratic differentials,
to the theory of logarithmic potentials and equilibrium measures.

Viewed in this light, the present paper explains that for the analysis
of $3 \times 3$ Riemann--Hilbert problems, the existence of a suitable
transformation is equivalent to the existence of an equation of the
form (\ref{eq:introcubic}) whose roots also satisfy a host of
properties.  For quartic potentials, we establish an existence theorem
via algebraic geometry.

\section{large $n$ asymptotic analysis of the Riemann-Hilbert problem
  \ref{rhp:04} }
\label{IdealSitu}

This paper studies the large $n$ asymptotic behavior of the solution
to the Riemann--Hilbert problem \ref{rhp:04}, under the assumption
that the functions $V_{j}$ are entire.  Furthermore, we will
assume that the integers $n_{j}$ 
grow to $\infty$ with $n$, in such a way that
\begin{eqnarray}
\label{eq:3}
\frac{n_{j}}{n} \to x_{j}, \ \  j=1,2.
\end{eqnarray}

\subsection{First step in the asymptotic analysis}
In analogy with the more standard case of orthogonal polynomials (see,
for example, \cite{op1}-\cite{op2}), we seek a vector of
functions 
\begin{eqnarray}
  \label{eq:4}
{\bf G}(z) = (g_{0}(z), g_{1}(z), g_{2}(z))  
\end{eqnarray}
satisfying the following set of assumptions.
\begin{enumerate}
\item There are $2$ disjoint sets, $I_{1}$ and $I_{2}$, each a compact subset of $\mathbb{R}$,with $\delta_{j}$ denoting the supremum of
  each set.
\item For each $j=1,2$, the function $g_{j}$ is analytic in
      $\mathbb{C}\setminus (-\infty, \delta_{j})$.
\item For $j=1,2$, each function $g_{j}$ behaves as follows at $\infty$:
  \begin{eqnarray}
    \label{eq:5}
    g_{j}(z) = \log{z} + \mathcal{O}\left(z^{-1}\right).
  \end{eqnarray}
\item The functions $g_{j}$ satisfy
  \begin{eqnarray}
    \label{eq:5a}
g_{0} = g_{1} + g_{2}.
  \end{eqnarray}
\end{enumerate}
In addition to these basic conditions, there will be a collection of
important conditions on these functions, collectively referred to as
the ``Ideal Situation''.  The goal is to deduce that there exists a
unique vector ${\bf 
G}$ that can 
be used to transform the Riemann-Hilbert problem
\ref{rhp:04} to a new Riemann-Hilbert problem in a form suitable for
asymptotic analysis.  To that end, the transformation that we seek is
as follows:

\begin{eqnarray}
  \label{eq:6}
&&  B := \diag\left( e^{- n \hat{\ell}_{0}/2},
e^{n_{1}\hat{\ell}_{1}/2},e^{n_{2} \hat{\ell}_{2}/2} \right) A(z) 
\diag\left( e^{- n(g_{0} - \hat{\ell}_{0}/2)},
e^{n_{1}(g_{1} - \hat{\ell}_{1}/2)}, e^{n_{2} (g_{2} - \hat{\ell}_{2}/2)}\right),
\end{eqnarray}
where ${\bf \hat{\ell}} = (\hat{\ell}_{0}, -\hat{\ell}_{1},
-\hat{\ell}_{2})$ is a vector of constants, which will be chosen along
with the vector ${\bf G}$.  For now, the only restriction on these
constants is that $n \hat{\ell}_{0} = n_{1} \hat{\ell}_{1} + n_{2}
\hat{\ell}_{2}$.  

The matrix $B$ inherits jump relationships across $\mathbb{R}$, that
are determined by the Riemann-Hilbert problem 
\ref{rhp:04} for $A$, along with the boundary value behavior of the
vector ${\bf G}$.  Therefore, $B$ satisfies the following
Riemann--Hilbert problem.

\begin{rhp}
\label{rhp:02}

\begin{enumerate}
\item[(a)] $B$ is analytic on $\mathbb{C} \setminus \mathbb{R}$. 
\item[(b)] The boundary values of $B$ satisfy 
\begin{equation} \label{RHgen2}
    B_+(x) = B_-(x)
    \begin{pmatrix}
    e^{-n (g_{0}^{+}-g_{0}^{-})} & e^{n g_{0}^{-} +
    n_{1} g_{1}^{+}- n V_{1}-n_{1}\ell_{1}} & 
e^{n g_{0}^{-} + n_{2} g_{2}^{+}- n V_{2} - n_{2}\ell_{2}} \\
    0 & e^{n_{1}(g_{1}^{+}-g_{1}^{-})}      &  0 \\
    0     & 0     & e^{n_{2}(g_{2}^{+}-g_{2}^{-})}
    \end{pmatrix},
\end{equation}
\item[(c)] As $z \to \infty$, we have
\begin{equation} \label{asympY}
    B(z) = \left(I + O\left(\frac{1}{z}\right)\right).
\end{equation}
\end{enumerate}

\end{rhp}
Note that for ease of notation, we have introduced the
constants $\ell_{j}$, $j= 1, 2$ (without hats).  These are defined
by $\ell_{1} = \frac{n \hat{\ell}_{0} + n_{1} \hat{\ell}_{1}}{2
  n_{1}}$, and $\ell_{2} = \frac{ n \hat{\ell}_{0} + n_{2} \hat{\ell}_{2}}{2 n_{2}}$.
\medskip

{\bf Ideal Situation}

In analogy with the case of orthogonal polynomials, a detailed
asymptotic analysis would be possible if the following were true.
\begin{itemize}
\item[A.] (As mentioned before) There are $2$ disjoint sets, $I_{1}$
  and $I_{2}$, each a compact subset of $\mathbb{R}$, so that for each
  $j=1,2$, $g_{j}$ is analytic on ${\mathbb{C}}\setminus (-\infty,
  \delta_{j})$ (where $\delta_{j}$ is the supremum of the set
  $I_{j}$).
\item[B.]  The function $g_{1}$ and its derivative $g_{1}'$ are 
continuous in the closure of 
$\mathbb{C}\setminus (-\infty, \delta_{1})$, and the functions $g_{2}$
and $g_{2}'$ are continuous in the closure of 
$\mathbb{C}\setminus (-\infty, \delta_{2})$.  That is,
$g_{j}$ and $g_{j}'$ achieve their boundary 
values in the sense of continuous functions.
\item[C.] For $j=1,2$, we have that for $z \in I_{j}$, the following
  two properties hold true:
(i) $\frac{1}{i}\left( g_{j}^{+} -  g_{j}^{-}\right)$ is real, and
  decreasing as one traverses $I_{j}$ according to its orientation,
  and (ii) 
  \begin{eqnarray}
    \label{eq:7}
    n g_{0}^{-} + n_{j} g_{j}^{+}- n V_{j} - n_{j} \ell_{j} =
    0.
  \end{eqnarray}
\item[D.] For $j=1,2$, we have that for $z \in \mathbb{R} \setminus
  I_{j}$, the following two properties hold true:
(i) $\frac{1}{i}\left( g_{j}^{+} - g_{j}^{-}\right)$ is
  constant, and (ii)
\begin{eqnarray}
  \label{eq:8}
\mbox{Re} \left[  n g_{0}^{-} + n_{j} g_{j} - n V_{j} - n_{j}
\ell_{j} \right] < 0.
\end{eqnarray}

\end{itemize}

{\bf Remark:}  The reader may wonder why conditions A-D above have been
described as an ideal situation.  The explanation requires
considerable calculation, but the basic idea is this:  on each subinterval 
$I_{j}$ of $\mathbb{R}$, the jump matrix may be factored in a manner
quite analogous 
to the matrix factorization that has been used for asymptotic analysis
of $2 \times 2$ Riemann-Hilbert problems in approximation theory over
the past 8 years.  The theory for the case of $2 \times 2$ matrices
can be essentially embedded in the more general setting.  The 
real evidence that this works is in the papers \cite{KWV},
\cite{BK2}, \cite{ABK}, and \cite{BK4}, where the authors prove in
special cases that 
the conditions A-D can be achieved, and then proceed to use this
information to compute a complete asymptotic description for the
Riemann-Hilbert problem.  

In Section \ref{sect:rhpanalysis}, we will carry out this
Riemann--Hilbert analysis, 
for the ``one-cut'' case.  Indeed, the reader will see that in this
case, the subsequent Riemann--Hilbert analysis is virtually identical to the
analysis carried out by Bleher and Kuijlaars \cite{BK2}.

{\bf Remark:}  The ``ideal situation'' described in A-D above does not
always hold true.  Even in the Gaussian case, for $a<1$, it turns out
that the desired sets $I_{1}$ and $I_{2}$ are not subsets of
$\mathbb{R}$, but rather they are piecewise analytic arcs (actually
each is a union of two line segments).  The more general setting can
be described as follows

\begin{itemize}
\item The Riemann--Hilbert problem \ref{rhp:04} is modified so that
  the jump is not across the real axis, but rather is across two
  contours, $\Gamma_{1}$ and $\Gamma_{2}$.  

\item  The ideal
situation would be if the sets $I_{j}$, no longer subintervals of $\mathbb{R}$,
become unions of piecewise analytic arcs, subsets of the contours
$\Gamma_{1}$ and $\Gamma_{2}$, so that the conditions described in A-D
above are true, on the sets $I_{j}$.  (The conditions in B must be
generalized so that the functions $g_{j}$ are analytic with
continuously differentiable  boundary values off semi-infinite subsets
of the contours $\Gamma_{1}$ and $\Gamma_{2}$.)
\end{itemize}

For $V$ convex and $a$ sufficiently large, it seems to be the case
that the more down-to-earth conditions of A-D, with most action taking
place on the real axis, should capture the situation, but this remains
to be seen.  On the other hand, when the more general setting does
occur, the existence of an algebraic curve may be deduced as well, and
the actual form of the equation depends only on the external field
$V$, and the parameters $a$, $n_{1}$ and $n_{2}$.
\vskip 0.3 in

Using the fact that $   n g_{0}=
n_{1}g_{1} + n_{2} g_{2} $, we may rewrite
the condition (\ref{eq:7}) as
\begin{eqnarray}
  \label{eq:10}
n_{j}\left( g_{j}^{-} + g_{j}^{+} \right) +  n_{k} g_{k}
- n V_{j} - n_{j} \ell_{j} =     0, \ \ \ z \in I_{j}, \ \ k,j = 1,2,
\ \ k \neq j.
\end{eqnarray}

In the next Section we will show that the
collection of boundary relations contained in (\ref{eq:10}) are
intimately related  to an equation of the form 
\begin{eqnarray}
  \label{eq:9}
  w^{3} + a_{2}(z) w^{2} + a_{1}(z) w + a_{0}(z) = 0
\end{eqnarray} 
in which the coefficients $a_{j}(z)$ are analytic for all $z \in
{\mathbb C}$.  We will show that the roots of this polynomial are
expressible explicitly in terms of the $g_{j}$'s and $V_{j}$'s.

\medskip
{\bf Remark:} For the case of $p\times p$ Riemann--Hilbert problems
associated to 
Random Matrices with Source, the collection of boundary relations
analogous to (\ref{eq:10}) should be related to an equation of the form 
\begin{eqnarray}
  \label{eq:9a}
  w^{p} + a_{p-1}(z) w^{p-1} + \cdots + a_{0}(z) = 0.
\end{eqnarray} 

\section{An algebraic curve}
\label{AlgCurve}

We begin by defining $f_{1}$ and $f_{2}$ as
follows:
\begin{eqnarray}
  \label{eq:11}
  f_{j}(z) = \frac{n_{j}}{n}g_{j}'(z), \ \ \ j = 1, 2.
\end{eqnarray}

The boundary relations (\ref{eq:10}) yield, upon differentiation, that
$f_{1}$ and $f_{2}$ satisfy 
\begin{eqnarray}
  \label{eq:11a}
& &   \left( f_{1}^{-} + f_{1}^{+} \right) + f_{2}
- V_{1}' =     0, \ \ \ z \in I_{1}, \\
\label{eq:11b}
& &   \left( f_{2}^{-} + f_{2}^{+} \right) + f_{1}
- V_{2}' =     0, \ \ \ z \in I_{2}.
\end{eqnarray}
In addition, if the condition that $\frac{1}{i}\left( g_{j}^{+} -
  g_{j}^{-}\right)$ is constant on $\mathbb{R} \setminus I_{j}$ holds
true (see
part D of the ``ideal situation'' of the previous section)
then clearly for each $j=1,2$, $f_{j}$ is analytic in $\mathbb{R}
\setminus I_{j}$.

\begin{lemma}\label{lem:1}
Assuming the hypotheses A - D set forth in the ``ideal situation''
of the previous section, we conclude that the functions
$f_{j}$ satisfy the following relations for all $z \in {\mathbb C}$.

\begin{eqnarray}
  \label{eq:12}
  & & f_{1}(z)^{2} + f_{1}(z) f_{2}(z) + f_{2}(z)^{2} -
  f_{1}(z)V_{1}'(z) - f_{2} (z)V_{2}'(z)
  = {\mathcal A}(z) \\
\label{eq:12a}
  & &   f_{1}(z) f_{2}(z) \left( f_{1}(z) + f_{2}(z)\right) -
  V_{1}'(z) f_{1}(z) \left(V_{1}'(z) - f_{1}(z) \right) - 
  V_{2}'(z) f_{2}(z) \left(V_{2}'(z) - f_{2}(z) \right) = {\mathcal B}(z) ,
\end{eqnarray}
where ${\mathcal A}(z)$ and ${\mathcal B}(z)$ are entire functions,
implicitly defined as follows:
\begin{eqnarray}
  \label{eq:13}
& & {\mathcal A}(z) = \frac{1}{2 \pi i} \int_{\gamma} \frac{f_{1}(s)
V_{1}'(s) + f_{2}(s) V_{2}'(s) }{s-z} ds \\ 
\label{eq:14}
& & 
{\mathcal B}(z) = 
\frac{1}{2 \pi i} \int_{\gamma} \frac{V_{1}'(s) f_{1}(s)
\left(V_{1}'(s) - f_{1}(s) \right) +  
  V_{2}'(s) f_{2}(s) \left(V_{2}'(s) - f_{2}(s) \right)}{s-z} ds
\end{eqnarray}
and in both (\ref{eq:13}) and (\ref{eq:14}), the contour of
integration $\gamma$ may be taken to consist of a finite union of simple closed Jordan
curves which do not intersect and whose interiors are disjoint, each
oriented in the 
clockwise direction, one subset of these contours encircling the
intervals comprising $I_{1}$, a second subset encircling the intervals
comprising 
$I_{2}$, and one final contour encircling the point $z$.  (Of course, since
$V_{1}$ and $V_{2}$ are entire, this may be deformed into one contour, 
a large circle oriented in the
clockwise direction, which encircles the sets $I_{1}$ and
$I_{2}$, and also encircles the point $z$.)

\end{lemma}

{\bf Proof}

Let $H_{2} = f_{1}(z)^{2} + f_{1}(z) f_{2}(z) + f_{2}(z)^{2} -
f_{1}(z)V_{1}'(z) - f_{2}(z) V_{2}'(z)$.  As defined, $H_{2}$ is
analytic in ${\mathbb C} 
\setminus \left( I_{1} \cup I_{2} \right)$.  However, the jump of $H_{2}$
across either of these sets may be computed.  Indeed, across $I_{1}$,
we have 
\begin{eqnarray}
  \label{eq:15}
  H_{2}^{+}(z) - H_{2}^{-}(z) = \left( f_{1}^{+}(z) - f_{1}^{-}(z)\right) \left\{  
f_{1}^{+}(z) + f_{1}^{-}(z) + f_{2}(z) - V_{1}'(z)
\right\} \ \ = \ \ 0.
\end{eqnarray}
where in the last equality, we have used the boundary relation
(\ref{eq:11a}).  In exactly the same way one may verify that $H_{2}$ also
possesses no jump across $I_{2}$.  Since we know that $f_{1}$
and $f_{2}$ are bounded, $H_{2}$ can 
have no isolated singularities, and so Morera's theorem tells us that
$H_{2}$ is in fact entire.  Equation (\ref{eq:12}) with ${\mathcal A}(z)$
defined in (\ref{eq:13}) follows by Cauchy's theorem.

The proof of (\ref{eq:12a}) goes along the same lines.  One begins by
defining 
\begin{eqnarray}
  \label{eq:16}
  H_{3} = f_{1}(z) f_{2}(z) \left( f_{1}(z) + f_{2}(z)\right) -
  V_{1}'(z) f_{1}(z) \left(V_{1}'(z) - f_{1}(z) \right) - 
  V_{2}'(z) f_{2}(z) \left(V_{2}'(z) - f_{2}(z) \right),
\end{eqnarray}
and observing that $H_{3}$ also possesses no jumps across $I_{1}\cup
I_{2}$, and hence it must be entire.  Then (\ref{eq:12a}) with
${\mathcal B}(z)$ defined in (\ref{eq:14}) follows by Cauchy's
theorem.  $\Box$

{\bf Remark}: The reader will note that the assumption that $V_{1}$
and $V_{2}$ should be entire is for comfort;  for example, if 
$V_{1}'$ and $V_{2}'$ are meromorphic functions, then the only
modification to the theorem is that ${\mathcal A}(z)$ and
$\mathcal{B}(z)$ are also meromorphic, as they might have poles at
those points that are poles of $V_{1}'$ and $V_{2}'$.  The integral
representations (\ref{eq:13}) and (\ref{eq:14}) for ${\mathcal A}$ and
${\mathcal B}$ remain true, with the contour of integration taken 
to consist of 3 simple closed Jordan
curves that do not intersect each other, each oriented in the
clockwise direction, one encircling $I_{1}$, the second encircling
$I_{2}$, and the third encircling the point $z$.  (It may not be
possible to deform these into a single contour without introducing
additional residues.)

{\bf Remark:}  In order to conclude that $H_{2}$ can have no isolated
singularities on $I_{1}$ or $I_{2}$, we don't actually need that the
boundary values be bounded.  The Lemma remains true if the
boundary values of $H_{2}$ lie in $L^{p}(I_{j})$ for some $1 < p <
2$.  This may be translated into conditions on the functions $f_{j}$,
but this is, at the present time, not particularly useful.

{\bf Remark:}  We note in passing that not all of the assumptions of
A-D are required for the proof of this Lemma, and in fact what is
actually required is A, B, (\ref{eq:7}), and item (i) of D.

The following uniqueness result follows immediately from Lemma \ref{lem:1}.

\begin{theorem}
\label{thm:1a}
Suppose that there are two disjoint sets, $I_{1}$ and
$I_{2}$, each one bounded and consisting of a disjoint union of
intervals.  Then the boundary value problem for
(\ref{eq:11a})-(\ref{eq:11b}) (with, for $j=1,2$, $f_{j}$ analytic in
$\mathbb{C}\setminus I_{j}$, and the boundary values of $f_{j}$
being bounded) possesses at most one solution.
\end{theorem}

{\bf Proof}
If there is another pair of functions $\hat{f}_{1}$ and $\hat{f}_{2}$
that satisfy  the same boundary value relations and asymptotics, then
their difference $F_{j} := f_{j} -\hat{f}_{j}$, $j=1,2$, satisfies the
following relations.
\begin{eqnarray}
  \label{eq:36}
& &   F_{1}^{+} + F_{1}^{-} + F_{2} = 0 , \ \ z \in I_{1}, \\
\label{eq:37}
& &   F_{2}^{+} + F_{2}^{-} + F_{1} = 0 , \ \ z \in I_{2}, \\
\label{eq:38}
& & F_{j} = \frac{c_{j}}{z^{2}}, z \to \infty.
\end{eqnarray}
Thus, Lemma \ref{lem:1} implies that for all $z \in \mathbb{C}$,
\begin{eqnarray}
  \label{eq:39}
& & F_{1}^{2} + F_{1} F_{2} + F_{2}^{2} = 0, \\  
\label{eq:40}
F_{1} F_{2} \left(  F_{1} + F_{2}  \right) = 0.
\end{eqnarray}
These identities finally imply that $F_{j} \equiv 0$ for $j = 1,2$,
and we have proven uniqueness.
$\Box$

\vskip 0.2in

Lemma \ref{lem:1} clearly demonstrates that the boundary value
relations (\ref{eq:11a})-(\ref{eq:11b}) are related to algebraic equations;
it is useful to recast these relations in terms of the following
result.

\begin{theorem}
\label{thm:1}

Define $r_{1}(z)$, $r_{2}(z)$, and $r_{3}(z)$ as follows:
\begin{eqnarray}
  \label{eq:17}
& &   r_{1} = f_{1} - \frac{1}{2} \left( V_{1}' -
V_{2}' \right)  \\
\label{eq:18}
& &   r_{2} = f_{2} + \frac{1}{2} \left( V_{1}' -
V_{2}' \right)  \\
\label{eq:19}
& & r_{3} =  - ( r_{1} + r_{2} ) + \frac{1}{2} \left( V_{1}' +
V_{2}'\right)  .
\end{eqnarray}
Then, under the assumptions of Lemma \ref{lem:1}, the following holds true:
\begin{eqnarray}
  \label{eq:20}
E(w,z) :=   \left(w - r_{1} \right)  \left(w - r_{2} \right)   \left(w - r_{3}
  \right) = w^{3} - {\mathcal C}_{2}(z) w^{2} +
  {\mathcal C}_{1}(z) w - {\mathcal C}_{0}(z), 
\end{eqnarray}
where 
\begin{eqnarray}
  \label{eq:21}
& & {\mathcal C}_{2}(z) = \frac{1}{2} ( V_{1}'(z) + V_{2}'(z)) \\
\label{eq:21b}
& &   {\mathcal C}_{1}(z) = \frac{-1}{4} \left(4{\mathcal A}(z) +
\left( V_{1}'(z) - V_{2}'(z)\right)^{2}   \right) \\
\label{eq:21c}
& & 
{\mathcal C}_{0}  = \frac{1}{8} \left[ 
4 {\mathcal A}(z) \left(V_{1}'(z) + V_{2}'(z)\right) - 8 {\mathcal B}(z)  - 
\left( V_{1}'(z) - V_{2}'(z)\right)^{2} \left( V_{1}'(z) + V_{2}'(z)\right)
\right].
\end{eqnarray}
\end{theorem}
{\bf Proof}

The proof is a straightforward exercise in algebra:  one must compute
the coefficients of  each power of $w$ in the expression for $E$,
using the definition (\ref{eq:17})-(\ref{eq:19}) of $r_{j}$, along
with (\ref{eq:12}) and (\ref{eq:12a}).  So, for example,
(\ref{eq:21}) may be verified as follows:
\begin{eqnarray}
  \label{eq:22}
{\mathcal C}_{2}(z) = r_{1} + r_{2} + r_{3} = \frac{1}{2} ( V_{1}'(z)
+ V_{2}'(z)), 
\end{eqnarray}
from (\ref{eq:19}).  The relations (\ref{eq:21b}) and
(\ref{eq:21c}) may be verified in a similar fashion.  $\Box$

{\bf Remark:}  As with Lemma \ref{lem:1}, the assumption that $V_{1}$
and $V_{2}$ should be entire is for convenience, and in particular if
$V_{1}$' and $V_{2}'$ are meromorphic, then the Theorem remains true
(but with ${\mathcal A}$ and ${\mathcal B}$ adjusted to admit the
possibility of poles at the pole locations of $V_{1}'$ and $V_{2}'$).

{\bf Remark:}  It is important to observe that Theorem \ref{thm:1}
yields nontrivial information about the functions $g_{1}$ and
$g_{2}$.  Indeed, the following result provides a drastic reduction in
the possible complexity of the sets $I_{1}$ and $I_{2}$.

\begin{theorem}
\label{thm:2}
Assuming the hypotheses A - D set forth in the ``ideal situation''
of the previous section,
each of the sets $I_{1}$ and $I_{2}$ consists of at most a finite union
of intervals. 
\end{theorem}
{\bf Proof}
Since $f_{1}$ is analytic in a neighborhood
of $\infty$, and since $I_{1}$ is a contour with
finite arclength, $f_{1}$ may be expressed as a Cauchy integral of its
jumps across $I_{1}$.  The same holds true for $f_{2}$, and so we have
\begin{eqnarray}
  \label{eq:23}
f_{j}(z) = \frac{1}{2 \pi i} \int_{I_{j}}
\frac{f_{j}^{+}(s)-f_{j}^{-}(s)}{s-z} ds, \ \ j = 1, 2.
\end{eqnarray}
On the other hand, the functions $r_{1}, r_{2},$ and $r_{3}$ are the
roots of a function $E(z,w)$ that is a cubic polynomial in $w$ with
coefficients that are entire functions of $z$, and so its roots are
analytic functions of $z$, with at worst a finite number of branching
points in any finite part of ${\mathbb C}$.  From the definition of
the $r_{j}$'s, we learn that the functions $f_{1}$ and $f_{2}$ can
have at most a finite number of branching points in the entire plane.
Each endpoint of a sub-arc of $I_{j}$ clearly corresponds to a
branching point, and so we conclude that $I_{1}$ and $I_{2}$ must each
consist of at most a finite union of intervals.
$\Box$

The conditions A through D above, when satisfied, should in principle
allow one to 
carry out a complete asymptotic analysis of the associated multiple
orthogonal polynomials, and obtain complete control of the asymptotic
statistics of the eigenvalues.  What has not been pinned down
completely under these conditions is the behavior near the
endpoints of $I_{1}$ and $I_{2}$.  This edge behavior depends on the
order of vanishing of the quantities $n g_{0}^{-} + n_{j} g_{j}^{+}- n
V_{j} - n_{j} \ell_{j}$ and $\frac{1}{i}\left( g_{j}^{+} -
  g_{j}^{-}\right)$ at the endpoints.  The  
following additional condition, expected to be the generic case,
implies that all edge behaviors are governed by the Airy equation.  We
will take $(\alpha,\beta)$ to be one of the intervals comprising the
set $I_{j}$ (which is now taken to consist of a finite union of
disjoint intervals).
\begin{itemize}
\item[E 1.]  The behavior near each left endpoint $\alpha$ is as
  follows.  There is $c_{\alpha}>0$  so that 
\begin{itemize}
\item[$\bullet$]
$g_{0}^{-}(z) + x_{j} g_{j}^{+}(z)-  V_{j}(z) - x_{j} \ell_{j} = -
c_{\alpha} |z - \alpha|^{3/2} + \mathcal{O}\left(|z-\alpha|^{5/2}\right)$ for $z$ near $\alpha$ and $z <
\alpha$.
\item[$\bullet$]
$\frac{1}{i}\left( g_{j}^{+}(z) -  g_{j}^{-}(z)\right) = c_{\alpha} | z -
\alpha|^{3/2} + \mathcal{O}\left(|z-\alpha|^{5/2}\right)$ for $z$ near $\alpha$ and $z > \alpha$.
\end{itemize}
\item[E 2.]  The behavior near each right endpoint $\beta$ is as
  follows.  There is $c_{\beta}>0$ so that  
\begin{itemize}
\item[$\bullet$]
$g_{0}^{-}(z) + x_{j} g_{j}^{+}(z)-  V_{j}(z) - x_{j} \ell_{j} = -
c_{\beta} |z - \beta|^{3/2} + \mathcal{O}\left(|z-\beta|^{5/2}\right)$ for $z$ near $\beta$ and $z >
\alpha$.
\item[$\bullet$]
$\frac{1}{i}\left( g_{j}^{+}(z) -  g_{j}^{-}(z)\right) = c_{\beta} | z -
\beta|^{3/2} + \mathcal{O}\left(|z-\beta|^{5/2}\right)$ for $z$ near $\beta$ and $z < \beta$.
\end{itemize}
\end{itemize}

\vskip 0.2in

\section{Polynomial external fields}
\label{PolyField}

According to the notation and definitions of Sections 2 and 3, we have
$V_{1}'=V'(x) + a$ and $V_{2}'=V'(x)-a$, and hence one may verify that
the desired algebraic curve has the form 
\begin{eqnarray}
  \label{eq:34}
  w^{3} - \left( V'(z) \right) w^{2} - \left[  {\mathcal
  A} + a^{2} \right] w - \left[ 
\left( {\mathcal A} -a^{2} \right) V'(z) -  {\mathcal B} \right] = 0.
\end{eqnarray}
The quantities ${\mathcal A}$ and ${\mathcal B}$ are entire functions,
as yet undetermined (aside from the fact that they satisfy
(\ref{eq:13}) and (\ref{eq:14})).  

Now if it is the case that $V(x)$ is a polynomial, more information
can be obtained about  ${\mathcal A}$ and ${\mathcal B}$.  In
this case, they satisfy
\begin{eqnarray}
  \label{eq:31}
&&  {\mathcal A}(z) = \mbox{Poly}\left[  
 -  V'(z) \left( f_{1}(z)+ f_{2} (z)\right)
\right], \\
&&
{\mathcal B}(z) = \mbox{Poly} \left[
-\left( V'(z) \right)^{2} \left(f_{1}(z) + f_{2}(z)\right)  + V'(z) \left(
f_{1}^2 + 2 a \left( f_{2}-f_{1} \right) + f_{2}^2\right)
\right]
\end{eqnarray}
where $\mbox{Poly}(Q)$ refers to the polynomial part of the Laurent
expansion of $Q$ valid for $z$ near $\infty$.
Clearly then, ${\mathcal A}$ is a polynomial of degree
$\mbox{deg}(V)-2$, and ${\mathcal B}$ is  a polynomial of degree $2
\mbox{deg}(V) - 3$.  So, if $V$ is a polynomial, then these
considerations determine the algebraic curve in question, up to the
choice of at most $3 \mbox{deg}(V) - 5$ undetermined coefficients.

Amazingly, the algebraic curve (\ref{eq:34}) is actually of much lower
degree than might be expected.  Indeed, some algebra shows that the
curve has the simplified form
\begin{eqnarray}
  \label{eq:34a}
  w^{3} - \left( V'(z) \right) w^{2}  +\mathcal{C}_{1}(z) \ w - \mathcal{C}_{0} = 0, 
\end{eqnarray}
where $
\mathcal{C}_{1}(z) =  \mbox{Poly}\left[  
V'(z) \left( f_{1}(z)+ f_{2} (z)\right)
\right] - a^{2} $ is a polynomial of degree $\mbox{deg}(V)-2$, and $C_{0}$
is a polynomial of degree $\mbox{deg}(V) - 1$.  To see that this
latter claim is true, note that one has the representation
\begin{eqnarray}
\label{eq:34b}
& &\mathcal{C}_{0}(z) = - a^{2} V'(z)  -
\mbox{Poly}\left[ V'(z)  \left(  f_{1}^{2}+ 2 a \left( f_{2}- f_{1}
    \right) + f_{2}^{2}\right) \right] + \\
& & \ \ \ \ \ \ \ \ \ \ \ \ \ \ \ \ \ \ \ \ + \ \mbox{Poly}\left[\left(  V'(z)
  \right)^{2} \left(  f_{1}(z) + f_{2}(z)\right) \right] 
-   V'(z)
\mbox{Poly}\left[V'(z)  \left(  f_{1}(z) + f_{2}(z)\right) \right],
\end{eqnarray}
which is easily seen to be a polynomial of degree $\mbox{deg}(V) - 1$.

\section{The existence problem.}

We have shown that {\it if there exists} sets $I_{1}$ and $I_{2}$ and
functions $f_{1}$ and $f_{2}$ satisfying the conditions of Lemma
\ref{lem:1}, then the functions $f_{1}$ and $f_{2}$ are related to the
roots of a particular algebraic curve.  On the other hand, for the
purposes of asymptotic analysis of Riemann--Hilbert problems, what is really
required is the {\it existence} itself:  given external fields $V_{1}$
and $V_{2}$, one needs to know that
functions $f_{1}$ and $f_{2}$, together with sets $I_{1}$ and $I_{2}$,
exist so that the original Riemann--Hilbert problem \ref{rhp:04} may
be ``regularized'' into the form of Riemann--Hilbert problem
\ref{rhp:02}, and subsequent asymptotic analysis may be carried out. 

If one knew the algebraic curve explicitly, then by careful analysis
one could prove that (i) sets $I_{1}$ and $I_{2}$ exist (they are the
branch cuts of the algebraic functions determined by the algebraic
equation), (ii) functions $f_{1}$ and $f_{2}$ exist (they are the
actual algebraic functions) and (iii) all the required reality, and
positivity/negativity conditions are satisfied (this is 
actually the most difficult part of the analysis).

Unfortunately, except in special Gaussian cases (see the next Section), the functions
$\mathcal{C}_{1}$ and $\mathcal{C}_{2}$ are not
completely determined by the considerations leading to, nor by the
proof of, Theorem \ref{thm:1}.  For example, if $V_{1}$ and $V_{2}$
are polynomials, then one can prove (again, see the
subsequent Section for concrete examples) that $\mathcal{C}_{1}$ and
$\mathcal{C}_{2}$  are polynomials in $z$, whose leading coefficients
are explicitly known.  However, the explicit determination of the
lower order terms does not follow from these considerations.  

In other words, one has not shown existence of sets $I_{j}$ and
functions $f_{j}$ unless one can prove that within the family of curves
parametrized by the coefficients left undetermined by Theorem
\ref{thm:1}, there exists an algebraic curve which satisfies all the
necessary conditions.  We will formulate this as an existence problem.

\begin{problem}Given polynomial (or more general) external fields
  $V_{1}$ and $V_{2}$, prove that there is an algebraic curve of the form 
(\ref{eq:20}) whose roots yield functions $f_{1}$ and $f_{2}$ as in
(\ref{eq:17})-(\ref{eq:18}) so that items A-D of the ``ideal
situation'' described near the end of Section 2 (or their
generalization to the case where the curves in question are not
necessarily real) are satisfied.
\end{problem}

\section{Existence theorems for quadratic and quartic external fields}
\label{ExistQuad}

\vskip 0.3in \noindent
{\bf The Gaussian case.}

The Gaussian case corresponds to $V=x^{2}/2$.  In this case, one may
verify that
\begin{eqnarray}
  \label{eq:28}
& &  {\mathcal A}(z) = -1, \\
& & {\mathcal B}(z) = - z.
\end{eqnarray}
And so, we see that the desired algebraic equation is 
\begin{eqnarray}
  \label{eq:29}
  w^{3} - z w^{2} -\left( a^{2} -1\right) w +
 a^{2} z + a \left( 2 x_{2} - 1\right)=0.
\end{eqnarray}
We observe that in the Gaussian case, there are no undetermined
coefficients at all, and, it is straightforward to verify (see, for
example, \cite{BK2}, \cite{ABK}, for the case $x_{1}=x_{2}=1/2$) that
there is $A$ so that for $a>A$, (\ref{eq:29}) possesses roots that
yield functions $g_{j}$ and sets $I_{j}$ so that the conditions A-D
hold true.  (In the case $x_{1} = x_{2}=1/2$, it is also known that
for $a<A$, the equation (\ref{eq:29}) determines two
piecewise analytic arcs as described in the third remark following the
description of the conditions A-D in Section \ref{AlgCurve} \cite{BK2,
ABK}).

\vskip 0.2in \noindent
{\bf The quartic case, with $x_{1}=x_{2}=1/2$.}
The quartic case corresponds to  $V(x) = x^{4}/4$.  In
this case, one may verify that
\begin{eqnarray}
& &  \mathcal{C}_{1} = z^{2} + c_{1} z + c_{0}, \\
& & \mathcal{C}_{0}= - a^{2} z^{3} + \hat{c}_{2} z^{2} + \hat{c}_{1} z + \hat{c}_{0}.
\end{eqnarray}
If we consider the special symmetric case in which $x_{1}=x_{2}=1/2$,
then there is an additional symmetry satisfied by the functions
$f_{1}$ and $f_{2}$, namely that
\begin{eqnarray*}
  f_{1}(z) = -f_{2}(-z).
\end{eqnarray*}
(This can be seen by considering $F_{1}(z) = -f_{2}(-z)$ and $F_{2}(z)
= -f_{1}(-z)$, showing that $F_{1}$ and $F_{2}$ satisfy the same
boundary value relations (\ref{eq:11a})-(\ref{eq:11b}) as $f_{1}$ and
$f_{2}$, and then using Theorem \ref{thm:1a}.)

This symmetry in turn implies that $\mathcal{C}_{0}$ is an odd
polynomial, and $\mathcal{C}_{1}$ is an even polynomial, and hence the
algebraic curve becomes
\begin{eqnarray}
\label{quart:01}
  w^{3} - z^{3} w^{2} + \left(z^{2} + \alpha \right)   w + a^{2} z^{3} +
  \beta z= 0, 
\end{eqnarray}
and there are only two undetermined coefficients, $\alpha$ and $\beta$.

In order to arrive at a problem of the form of Riemann--Hilbert
problem \ref{rhp:02}, the existence of sets $I_{1}$ and $I_{2}$, along
with functions $f_{1}$ and $f_{2}$ must be established, as described
in Section 3.  As a demonstration that this is possible, we have the
following result.

\begin{theorem}
\label{thm:3}
There is $A>0$ so that for all $a> A$, there exists coefficients
$\alpha(a)$ and $\beta(a)$ so that the following properties,
pertaining to  equation (\ref{quart:01}) with
$\alpha = \alpha(a)$ and $\beta = \beta(a)$, are true.
\begin{enumerate}
\item  The three solutions
$\{r_{j}(z) \}_{j=1}^{3}$ of 
the equation (\ref{quart:01}), determined uniquely as
algebraic functions of $z$ for $z \in \mathbb{C}$ that possess
standard meromorphic Laurent expansions at $\infty$, satisfy the
following asymptotic behavior for $z \to \infty$:
\begin{eqnarray}
\label{RMT:root3}
&&r_{3} = z^{3}  - \frac{1}{z} + \mathcal{O} \left(z^{-2} \right), \\
\label{RMT:root2}
&&r_{2} = a + \frac{1}{2 z} + \mathcal{O} \left(z^{-2} \right), \\
\label{RMT:root1}
&&r_{1} = -a + \frac{1}{2 z} + \mathcal{O} \left(z^{-2} \right).
\end{eqnarray}
\item
The functions $f_{1}$ and $f_{2}$, defined by
\begin{eqnarray}
f_{1}(z) = r_{1}(z) + a, \ \ \ \ \  f_{2}(z) = r_{2}(z) - a,
\end{eqnarray}
possess the following asymptotic expansions for $z \to \infty$:
\begin{eqnarray}
\label{RMT:fexp}
f_{j} = \frac{1}{2 z} + \mathcal{O}\left( z^{-2} \right), \ \ \   j = 1,2.
\end{eqnarray}
\item
There are two real positive real numbers $\gamma_{1} = \gamma_{1}(a)$
and $\gamma_{2} = \gamma_{2}(a)$ so that the function $f_{2}$ may be
taken to be analytic in $\mathbb{C} \setminus I_{2}$ where
$I_{2} =[\gamma_{1} , \gamma_{2}]$.
\item
The function $f_{1}$ may be taken to be analytic in $\mathbb{C}
\setminus I_{1}$ where 
$I_{1} =[-\gamma_{2} , -\gamma_{1}]$.
\item
The following boundary relations hold true:
\begin{eqnarray}
\label{eqn:BVRel1}
& &   \left( f_{1}^{-} + f_{1}^{+} \right) + f_{2}
- V_{1}' =     0, \ \ \ z \in I_{1}, \\
\label{eqn:BVRel2}
& &   \left( f_{2}^{-} + f_{2}^{+} \right) + f_{1}
- V_{2}' =     0, \ \ \ z \in I_{2}.
\end{eqnarray}
\item For $z \in I_{j}$, we have
\begin{eqnarray}
\label{eqn:MeasReal}
\frac{1}{i}\left( f_{j}^{+}(z) - f_{j}^{-}(z)\right) <0.
\end{eqnarray}
\item For $z \in \mathbb{R} \setminus I_{1}$, we have
\begin{eqnarray}
\label{eqn:VariationIneq}
 \int_{-\gamma_{2}}^{z} \mbox{Re } \left( f_{1}^{-}(s) + f_{1}^{+}(s) + f_{2}^{-}(s)
- V_{1}'(s) \right) \  ds  < 0,
\end{eqnarray}
and for $z \in \mathbb{R} \setminus I_{2}$, we have the analogous
inequality:
\begin{eqnarray}
\label{eqn:VariationIneq2}
\int_{\gamma_{2}}^{z} \mbox{Re } \left( f_{2}^{-}(s) + f_{2}^{+}(s) + f_{1}^{-}(s)
- V_{2}'(s) \right) \ ds  < 0,
\end{eqnarray}
\end{enumerate}
\end{theorem}
{\bf Remark:}  We note that for the quartic case
considered here, Theorem \ref{thm:3} solves the Existence
Problem as described in Section 3, for all $ a > A$.

{\bf Proof of (1) and (2):}  Straightforward asymptotic calculations for $z\to\infty$
(which are left to the diligent reader)
show that the three roots $\{r_{j} \}_{j=1}^{3}$ have
Laurent expansions of the form (\ref{RMT:root3})-(\ref{RMT:root1}),
and it is immediately clear that (\ref{RMT:fexp}) holds true.

\bigskip

{\bf  Proof of (3) and (4):  }Whether one calculates the resultant of
the polynomial  
\begin{eqnarray*}
p_{3}(z,w) =   w^{3} - z^{3} w^{2} + \left(z^{2} + \alpha \right)   w
+ a^{2} z^{3} + \beta z
\end{eqnarray*}
with
its derivative $(\partial / \partial w) p_{3}(z,w)$ or just solves the equation
(\ref{quart:01}) directly, one arrives at the following 12th degree polynomial
in the variable $z$ whose roots are potential branch points for the
roots $\{r_{j}(z)\}_{j=1}^{3}$:
\begin{eqnarray}
& & \tilde{q}(z; \alpha, \beta, a) = 36 a^2 z^{12}+9 (4 \beta +1) z^{10}+9 \left(2 \alpha -18
   a^2\right) z^8+9 \left(-27 a^4-18 \alpha  a^2+\alpha
   ^2-18 \beta -4\right) z^6+\\
\nonumber
& & \ \ \ \ \ + 9 \left(-54 \beta  a^2-12
   \alpha -18 \alpha  \beta \right) z^4+9 \left(-12 \alpha
   ^2-27 \beta ^2\right) z^2-36 \alpha ^3.
\end{eqnarray}
The polynomial $\tilde{q}$ is even and calculations are greatly
simplified by considering 
\begin{eqnarray}
\label{eq:46}
& & q(t) = \tilde{q}(\sqrt{t}) =  36 a^2 t^{6}+9 (4 \beta +1) t^{5}+9 \left(2 \alpha -18
   a^2\right) t^{4}+9 \left(-27 a^4-18 \alpha  a^2+\alpha
   ^2-18 \beta -4\right) t^{3}+\\
\nonumber
& & \ \ \ \ \ \ \ \ \ \ \ \ \ \ \ \ \ \ \ \ \ \ \ + 9 \left(-54 \beta  a^2-12
   \alpha -18 \alpha  \beta \right) t^{2}+9 \left(-12 \alpha
   ^2-27 \beta ^2\right) t-36 \alpha ^3.
\end{eqnarray}
The goal is to find $(\alpha, \beta)$ so that this polynomial has
exactly two positive simple roots and two double roots (that will turn
out to be complex).  We consider the resultant of $q(t)$ with its
derivative $q'(t)$:
\begin{eqnarray}
\mbox{Resultant}\left[q(t), q'(t) \right] = B_{1}(\alpha, \beta, a)
B_{2} (\alpha, \beta, a),
\end{eqnarray}
where
\begin{eqnarray}
& & B_{1}(\alpha, \beta, a) = 729 \alpha  a^6+243 \left(\alpha ^2-3 \beta
\right) a^4+27 \alpha  \left(\alpha ^2-15 \beta -1\right) a^2+\alpha
^4+27 \beta  (3 \beta +1)^2-36 \alpha ^2 \beta, \\
& & 
B_{2}(\alpha, \beta, a) = 729 \alpha ^3 a^{10}+729 \left(3 \alpha
  ^4-\beta  \alpha ^2-\beta ^3\right) a^8+27 \left(81 \alpha ^5+(8-171
  \beta ) \alpha 
   ^3-72 \beta ^3 \alpha \right) a^6+\\
& & \nonumber \ \ \ \ \ \ \ \ \ \ \ \ \ \ \ \ \ \ \ \ \ \ + 27 \left(27 \alpha ^6-20 (9 \beta
   +1) \alpha ^4-2 \beta  \left(28 \beta ^2-47 \beta 
   +4\right) \alpha ^2+8 \beta ^3 (12 \beta -1)\right) a^4+\\
& & \nonumber \ \ \ \ \ \ \ \ \ \ \ \ \ \ \ \ \ \ \ \ \ \ \ \ \ + \left(-27
 (36 \beta +1) \alpha ^5+4 \left(216 \beta ^3+810 \beta 
   ^2+189 \beta +4\right) \alpha ^3+576 \beta ^3 (4 \beta +1) \alpha
\right) a^2+ \\
\nonumber & & \ \ \ \ \ \ \ \ \ \ \ \ \ \ \ \ \ \ \ \ \ \ \ \ \ \ \ \
\ -\beta  (4 \beta +1)^2 \left(-27 \alpha ^4+8 
   (9 \beta +2) \alpha ^2+16 \beta ^2 (4 \beta +1)\right).
\end{eqnarray}
\begin{lemma}
\label{lem:B2}
There is $A$ so that for all $a > A$, the following statement is true:
For real $\alpha$ and $\beta$, there is an isolated zero of
$B_{2}(\alpha, \beta)$, which is a local maximum, occurring at
$(\alpha^{*}(a), \beta^{*}(a))$.  The quantities $\alpha^{*}(a)$ and
$\beta^{*}(a)$ possess the following expansions:
\begin{eqnarray}
\label{eq:42}
& & \alpha^{*}(a) = a^{2} \left(  
-1 + a^{-4/3} + \left(\frac{1}{27}\right) a^{-12/3} + a^{-16/3} \
\sum_{j = 0}^{\infty} \alpha_{j} \left( a ^{-4/3} \right)^{j} \right)
\\
& & 
\label{eq:43}
\beta^{*}(a) = a^{4/3} \left(  
1 - \left( \frac{1}{3} \right)a^{-4/3} +  a^{-12/3} \ 
\sum_{j = 0}^{\infty} \beta_{j} \left( a ^{-4/3} \right)^{j} \right)
\end{eqnarray}
\end{lemma}
{\bf Proof}  

For the polynomial $B_{2}(\alpha, \beta)$ to have a local maximum with
value $0$, it must vanish together with its first partial
derivatives. To prove this, we first observe that 
\begin{itemize}
\item[i] the resultant of $B_{2}$ with $\partial_{\alpha} B_{2}$
  (eliminating $\beta$) coincides, up to a constant factor, with the
  resultant of $B_{2}$ with $\partial_{\beta} B_{2}$:
\begin{eqnarray}
\mbox{Resultant}\left[
    B_{2}, \partial_{\alpha} B_{2}\right] (\alpha) =
\frac{a^{4}}{2048}
\mbox{Resultant}\left[
    B_{2}, \partial_{\beta} B_{2}\right] (\alpha).
\end{eqnarray}
\item[ii] Moreover, there is a unique value of $\alpha$ (called
  $\alpha_{*}$) with the
  asymptotic expansion (\ref{eq:42}) such that the above resultants
  vanish.  There are twelve other roots (of varying multiplicities),
  which remain bounded away from 
  this root by a distance of $\mathcal{O}\left(a^{2/3} \right)$.

\item[iii] In similar fashion, the resultants of $B_{2}$ with its first partials,
eliminating $\alpha$ instead of $\beta$, are also equivalent modulo a
constant factor:
\begin{eqnarray}
\mbox{Resultant}\left[
    B_{2}, \partial_{\alpha} B_{2}\right] (\beta) =
-a^{4}
\mbox{Resultant}\left[
    B_{2}, \partial_{\beta} B_{2}\right] (\beta). \ \ \ \ ).
\end{eqnarray}
\item[iv] Furthermore, there is a unique value of $\beta$ (called
  $\beta_{*}$) with the
  asymptotic expansion (\ref{eq:43}) so that {\it these} two
  resultants vanish.  There are 12 other roots (of varying
  multiplicities), which remain bounded 
  away from this root by a distance of $\mathcal{O}\left(
    a^{4/3}\right)$ (independent of $a$). 
\end{itemize}

The assertions ii and iv above may be verified by computing the large $a$
behavior of these resultants.  

Next one considers the Sylvester matrix associated to 
$\mbox{Resultant}\left[ B_{2}, \partial_{\beta} B_{2}\right](\alpha)$
(i.e. the matrix whose determinant is this resultant),
and shows that for $\alpha = \alpha_{*}$, there is an eigenvector of
the form 
\begin{eqnarray}
\left(
\beta(\alpha_{*})^{10}, 
\beta(\alpha_{*})^{9} ,
\dots, \beta(\alpha_{*}), 1
\right)^{t}
\end{eqnarray} with $\beta(\alpha_{*})$ possessing the first few terms of the
asymptotic expansion (\ref{eq:43}).
Since $\beta(\alpha_{*})$ must be a root of the polynomial
$\mbox{Resultant}\left[ B_{2}, \partial_{\beta} B_{2}\right](\beta)$,
it follows that $\beta(\alpha_{*}) = \beta_{*}$.

It now follows that $B_{2}$, along with its first partial derivatives,
vanishes for $(\alpha, \beta) = (\alpha_{*}, \beta_{*})$.

Again using the fact that $a$ may be taken large, one may verify that
for $(\alpha, \beta)$ near $(\alpha_{*}, \beta_{*})$, the surface
defined by $B_{2}(\alpha, \beta)$ satisfies
\begin{eqnarray}
\mbox{Hess}\left( B_{2}\right)(\alpha_{*}, \beta_{*}) = 
-4782969 \ a^{80/3}\left( 1 + \mathcal{O}\left( a^{-4/3}\right) \right).
\end{eqnarray}

That an $A$ exists so that for all $a>A$ the above
holds true is a straightforward exercise in perturbation theory.
$\Box$

\smallskip

{\bf Remark:}  From this point on, we shall take $a>A$, and $\alpha =
\alpha_{*}$, $\beta = \beta_{*}$.  However, to avoid cumbersome
notation, we will usually not emphasize this.  For example, $q(t)$
will now refer to $q(t; \alpha = \alpha_{*}, \beta = \beta_{*} )$,
etc.

\medskip

A consequence of Lemma  \ref{lem:B2} is that the polynomial $q(t)$ defined in
(\ref{eq:46}) possesses two complex double roots.  One sees this by
first carrying out an asymptotic expansion of $q(t)$ for $a \to
\infty$, using the asymptotic expansions (\ref{eq:42}) and
(\ref{eq:43}), to see that $q(t)$ possess the following asymptotic
expansion:
\begin{eqnarray}
& &  a^{-6} \ q(a^{2/3}\zeta) = 36 ( \zeta - 1)^{2} (\zeta^{2} + \zeta +
  1)^{2} +  36\,a^{-4/3}\, \left( 1 + \zeta + \zeta^2 \right) \,\left( -3 - 6\,\zeta^2
    + \zeta^3 \right)  + \\
\nonumber & & \ \ \ \ \ \ \ \ \ \ \ \ \ \ \ \ \ \ \ \ \ \ -3\,
a^{-8/3}\, \left( -36 + 9\,\zeta - 18\,\zeta^2 +  
    51\,\zeta^3 - 6\,\zeta^4 + \zeta^5 \right) + \\
\nonumber & & \ \ \ \ \ \ \ \ \ \ \ \ \ \ \ \ \ \ \ \ \ \ \ \ \ \ \ +
18\, a^{-12/3}\left( -2 + \zeta 
  \right) \,\left( 1 - \zeta + \zeta^2 \right) - 27 a^{-16/3} \zeta.
\end{eqnarray}
Perturbative calculations then show that for $a$ sufficiently large, $q(a^{2/3} \zeta)$ possesses
exactly two 
real roots.  These are simple and positive (near $\zeta = 1$).  In
fact, the reader may verify that the two roots, which we define to be 
$\gamma_{1}^{2}$ and $\gamma_{2}^{2}$ (where $\gamma_{1}$ and
$\gamma_{2}$ are two positive numbers) satisfy:
\begin{eqnarray}
\label{eq:rootAss2}
&&\gamma_{2}^{2} = a^{2/3}\left[ 1 + 2 \sqrt{\frac{2}{3}} a^{-2/3} +
  \cdots \right] \\
\label{eq:rootAss1}
&&\gamma_{1}^{2} = a^{2/3}\left[ 1 - 2 \sqrt{\frac{2}{3}} a^{-2/3} +
  \cdots \right]
\end{eqnarray}
Therefore $q(t)$ has 
four complex roots, appearing in complex conjugate pairs.  Since
$q(t)$ must have at least one double root, there must be two double
roots.  Therefore $q(t)$ factors as follows:
\begin{eqnarray}
q(t) = 36 a^{2} (t - \gamma_{1}(a)^{2})(t - \gamma_{2}(a)^{2}) \left[
  ( t - \lambda_{*})  ( t - \overline{\lambda_{*}}) \right]^{2},
\end{eqnarray}
with $ \mbox{Imag} \left( \lambda_{*} \right) > 0$.

\medskip

Now if one solves the cubic (\ref{quart:01}), one finds
\begin{eqnarray}
\label{eq:47}
w = \frac{1}{3} z^{3} + \ \left( \frac{- R(z) + 3^{-5/2}\sqrt{-q(z^{2})}}{2}
\right)^{1/3} +  
\frac{H(z) }{\left(
    \frac{- R(z) + 3^{-5/2}\sqrt{-q(z^{2})}}{2}  \right)^{1/3}} \ ,
\end{eqnarray}
where 
\begin{eqnarray}
& & R(z) = -\frac{2}{27} z^{9} + \frac{1}{3} z^{5} + \frac{\alpha}{3}
z^{3} + a^{2} z^{3} + \beta z\ , \\
& & H(z) = \frac{1}{9}z^6 - \frac{1}{3}z^2 - \frac{1}{3}\alpha \ ,
\end{eqnarray}
and the cube roots appearing in (\ref{eq:47}) are all the same choice
of branch, but not necessarily the principal branch (the three different
choices yield different solutions to the cubic (\ref{quart:01})).
Similarly, the square roots appearing in
(\ref{eq:47}) are all the same choice of branch, not necessarily the
principal branch.  (However, different choices of the branch of the
square root do not yield different solutions.)

\medskip

{\bf Remark:}  The representation (\ref{eq:47}), although concise, is
at present only useful for local analyses since the choice of
branches, usually a 
delicate matter, has been avoided.  We will use this representation
together with local analysis near zeros of $q(z^{2})$ to describe the
possible branch points of the solutions to (\ref{quart:01}), and later
we will make a particular choice of branches to carry out a global analysis.

\medskip

We already know that the only potential branching points where a
solution to (\ref{quart:01}) could fail to be analytic are the simple roots
of $q(z^{2})$.  Yet another asymptotic analysis for large $a$ shows
that $R(z)$ does not vanish at any of these roots, and so
we conclude that there are only four branch points, namely $\pm
\gamma_{1}(a)$ and $\pm \gamma_{2}(a)$, and at each of these points,
the solutions to (\ref{quart:01}) exhibit square-root singularities.
(So, for example, in a vicinity of $\gamma_{1}$, the solutions are
analytic functions of $\sqrt{z - \gamma_{1}}$.  )

At each of the four branch points, the square root appearing in
(\ref{eq:47}) vanishes, while $R(z)$ does not, and so a local analysis
in a vicinity of each branch point can be obtained by choosing a
particular branch for the square root (as will be done below), and
then, for each distinct choice of branch of the cube-root, use the
Taylor expansions
\begin{eqnarray}
  \label{eq:49}
& &  \left(- R + 3^{-5/2} \sqrt{-q(z^{2})} \right)^{1/3} = 
\sqrt[3]{\left| \frac{-R}{2} \right| }e^{i \phi/3}
\left\{
1 - \frac{1}{3}\left(\frac{3^{-5/2} \sqrt{-q(z^{2})}}{R(z)}\right) -
\frac{1}{9}\left(\frac{3^{-5/2} \sqrt{-q(z^{2})}}{R(z)}\right)^2 + \right. \\
\nonumber & & \hspace{2.4in} - \left.
\frac{5}{81}\,\left(\frac{3^{-5/2} \sqrt{-q(z^{2})}}{R(z)}\right)^3 -
\frac{10}{243}\,\left(\frac{3^{-5/2} \sqrt{-q(z^{2})}}{R(z)}\right)^4 + \ \cdots
\right\}, \\
  \label{eq:49b}
& &  \ \left(- R - 3^{-5/2} \sqrt{-q(z^{2})} \right)^{1/3} = 
\sqrt[3]{\left| \frac{-R}{2} \right| }e^{i \phi/3}
\left\{
1 + \frac{1}{3}\left(\frac{3^{-5/2} \sqrt{-q(z^{2})}}{R(z)}\right) -
\frac{1}{9}\left(\frac{3^{-5/2} \sqrt{-q(z^{2})}}{R(z)}\right)^2 + \right. \\
\nonumber & & \hspace{2.4in} + \left.
\frac{5}{81}\,\left(\frac{3^{-5/2} \sqrt{-q(z^{2})}}{R(z)}\right)^3 -
\frac{10}{243}\,\left(\frac{3^{-5/2} \sqrt{-q(z^{2})}}{R(z)}\right)^4 + \ \cdots
\right\},
\end{eqnarray}
where the phase $\phi = \phi(z)$ is the complex argument of $-R(z)$.
Different choices of this phase yield the three different solutions to
(\ref{quart:01}).

However, before doing this, it is advantageous to eliminate the
radical from the denominator of (\ref{eq:47}):
\begin{eqnarray}
& &   \label{eq:50}
  \frac{H(z) }{\left(
    \frac{- R(z) + 3^{-5/2} \sqrt{-q(z^{2})}}{2}  \right)^{1/3}} = 
e^{-2 i \phi/3}
\left(
\frac{-R(z) - 3^{-5/2} \sqrt{-q(z^{2})}}{2}
\right)^{1/3}.
\end{eqnarray}

Now using the convergent Taylor expansions (\ref{eq:49}) and
(\ref{eq:49b}) in (\ref{eq:47}), the reader may verify (with some
sweat) that exactly one of the three possible branches of the
cube-root is analytic in a neighborhood of each branch point.  Indeed
for that one choice of cube-root, all odd powers of the square-root
happen to vanish!  For example, for $z$ real and near $ \gamma_{2}$,
it turns out that $R(z) > 0$ (see Lemma \ref{lem:RMTaux} below) and
hence one must choose $\phi$ to be an odd multiple of $\pi$.  The
three choices $-\pi, \pi, 3 \pi$ will yield all possible solutions to
the cubic equation (\ref{quart:01}).  Now using the Taylor expansions
(\ref{eq:49}) and (\ref{eq:49b}), one finds
\begin{eqnarray}
\label{eq:51}
&& 
w(z) - \frac{1}{3} z^{3} = 
\left( \frac{- R(z) + 3^{-5/2} \sqrt{-q(z^{2})}}{2}
\right)^{1/3} +  
\frac{H(z) }{\left(
    \frac{- R(z) + 3^{-5/2} \sqrt{-q(z^{2})}}{2}  \right)^{1/3}} = \\
&&\nonumber \hspace{-0.2in}
\sqrt[3]{\left| \frac{-R}{2} \right|} \left( 
\sum_{j=0, \ j \mbox{ even}}^{\infty}2 c_{j} \left(
  \frac{3^{-5/2} \sqrt{-q(z^{2})}}{R(z)} \right)^{j} \cos{ (\phi/3)} \ - \ 
\sum_{j=1, \ j \mbox{ odd}}^{\infty}2 i \tilde{c}_{j} \left(
  \frac{3^{-5/2} \sqrt{-q(z^{2})}}{R(z)} \right)^{j} \sin{ (\phi/3)}  \right),
\end{eqnarray}
with $c_{0} = 1$, $\tilde{c}_{1}=1/3$, 
and clearly the choice $\phi=3 \pi$ is the only one for which all odd
powers of the square root vanish, yielding a solution (\ref{eq:47}) to
(\ref{quart:01}) which is analytic in a neighborhood of $\gamma_{2}$,
while the remaining two are not.

\medskip

{\bf Remark}:  Observe that for the other two choices of the angle
$\phi$, namely $-\pi $ and $\pi$, the two corresponding roots
coincide at $z = \gamma_{2}$.  These two roots are discontinuous
across $I_{2}  = (\gamma_{1}, \gamma_{2})$.  It follows that the
boundary value on one side of $I_{2}$ of one of these two roots must
coincide with the opposing boundary value of the {\it other} root.
Similarly, for the two roots that are not analytic across $I_{1}$,
their boundary values from opposing sides coincide, and at the
endpoints of the interval $I_{1}$, those two roots are equal.

\medskip

{\bf Global analyticity properties}

In order to describe the global analyticity properties of the
solutions to (\ref{quart:01}), we will make the following choice for
the branch of the square root appearing in (\ref{eq:47}):
\begin{eqnarray}
\label{eq:48}
\sqrt{-q(z^{2})}: = 6 i a  \left[
  ( z^{2} - \lambda_{*})  ( z^{2} - \overline{\lambda_{*}}) \right] \ 
\sqrt{z - \gamma_{1}(a)} \ \sqrt{z - \gamma_{2}(a)} \ \sqrt{z +
  \gamma_{1}(a)} \ \sqrt{z+ \gamma_{2}(a)} \  
\end{eqnarray}
where now $\sqrt{\cdot}$ is the principal branch.  The function
appearing in (\ref{eq:48}) is analytic in $\mathbb{C} \setminus \left\{
  [-\gamma_{2},-\gamma_{1}]  \cup [\gamma_{1}, \gamma_{2}] \right\}$.

Using this choice of branch for $\sqrt{-q(z^{2})}$ in (\ref{eq:47}),
along with the fact that $R(z)$ is strictly positive on $I_{2}$, it
follows that the one solution which is analytic in a neighborhood of
$\gamma_{2}$ can be continued to a fixed neighborhood of the
interval $[\gamma_{1}, \gamma_{2}]$.  This solution is then analytic
for $z \in\mathbb{C} \setminus  [-\gamma_{2}, -\gamma_{1}]$ (of
course, it cannot be entire, and so it must have a branch cut, which
can only be across $(-\gamma_{2}, -\gamma_{1})$).  This
solution shall be referred to as $r_{1}(z;a)$.

A similar argument shows that there is another solution, which will be
called $r_{2}(z;a)$, that is analytic in $\mathbb{C} 
\setminus [\gamma_{1}, \gamma_{2}]$.

{\bf Remark:} In order to complete the proof of (3) and (4) of Theorem
\ref{thm:3}, we will show that the solutions $r_{1}(z;a)$ and
$r_{2}(z;a)$ possess the $z \to \infty$ asymptotic expansions
(\ref{RMT:root1}) and (\ref{RMT:root2}), respectively.  We will only
consider the case of $r_{1}(z;a)$, as the proof that the solution
$r_{2}(z;a)$ possesses the expansion (\ref{RMT:root2}) is entirely
similar.

The starting point for this global analysis is the signature of the
functions $R(z)$, $H(z)$, and $q(z^{2})$ in a vicinity of
$\gamma_{2}(a)$, and we are once again aided by the fact that $a$ may
be taken large.  Indeed, we have the following lemma.
\begin{lemma}
\label{lem:RMTaux}
There is $A>0$ so that for all $a>A$, the following statements are
true.
\begin{itemize}
\item There is $\eta(a) > \gamma_{2}(a) $ so that the function $R(z)$
is positive on the interval $(0, \eta)$, and is negative
on $(\eta(a), \infty)$.
\item  The function $H(z)$ is strictly positive for all $z >
0$
\end{itemize}
\end{lemma}
{\bf Proof} 

Using the rescaling $z \mapsto a^{1/3} \zeta$, we find 
\begin{eqnarray}
  \label{eq:54}
& &a^{-3} R(a^{1/3} \zeta) = p(\zeta) + a^{-4/3} \hat{p}(\zeta), 
\end{eqnarray}
where
\begin{eqnarray}
\label{eq:55}
& & p(\zeta) = \frac{1}{27} {\zeta }^3\,\left( 27 + 9\,
  \left( \frac{\alpha_{*}}{a^{2}} \right)  - 2\,{\zeta }^6 \right) \\
\label{eq:56}
& &
\hat{p}(\zeta) = \left(  \left(\frac{\beta_{*}}{a^{4/3}} \right) \,\zeta  +
  \frac{{\zeta }^5}{3}\right).
\end{eqnarray}
Now $\alpha_{*}$ and $\beta_{*}$ possess asymptotic expansions
(cf. (\ref{eq:42}),(\ref{eq:43})) so that the coefficients
$\frac{\alpha_{*}}{a^{2}}$ and $\frac{\beta_{*}}{a^{4/3}}$ in
(\ref{eq:55}) and (\ref{eq:56}) are uniformly bounded, and 
\begin{eqnarray}
  \label{eq:53}
\frac{\alpha_{*}}{a^{2}} \rightarrow -1, \ \ \ \ \ \ \frac{\beta_{*}}{a^{4/3}}
\rightarrow 1.  
\end{eqnarray}
The reader may verify that the leading order term, $p(\zeta)$,
possesses one positive real root 
which approaches $3^{1/3}$.  Moreover, $p(\zeta)$ is positive to
the left of this root, and negative to the right.

Next we consider $\zeta $ in a fixed
interval of the form $[\frac{\gamma_{2}}{a^{1/3}}, X]$.  The 
function $\hat{p}(\zeta)$ is uniformly bounded on this interval, and
so the function $a^{-3} R(a^{1/3}\zeta)$ can only possess roots in a
vicinity of the root of $p(\zeta)$.  However, this is a simple root of
$p(\zeta)$, and the derivative $(d / d \zeta) a^{-3} R(a^{1/3}
\zeta)$ is strictly negative in a vicinity of this root, guaranteeing
that $a^{-3} R(a^{1/3} \zeta)$ possesses exactly one simple root in
the interval $[\frac{\gamma_{2}}{a^{1/3}}, X]$.

Lastly one considers the behavior for $\zeta$ large, where the
function $p(\zeta)$ dominates the function $\hat{p}(\zeta)$, and one
may then choose $X$ large enough (but fixed) so that $R(a^{1/3}
\zeta)$ possesses no roots beyond $X$, and this
finishes the proof of the first claim of the Lemma.

The second claim can be established by similar reasoning, using the
asymptotic representation
\begin{eqnarray}
\nonumber & &
a^{-2} H(a^{1/3} \zeta) =
\frac{1}{9} \left(  \zeta^{6} -
3\,\left(\frac{\alpha_{*}}{a^{2}} \right)  \right) -
a^{-4/3} \ \left( \frac{1}{3} \ \zeta^{2} \right)
\end{eqnarray}
It is straightforward to verify that the leading order term
on the right hand side is strictly positive for
$\zeta > \gamma_{2}/a^{1/3}$, and the remaining details of the proof
of the second claim are left to the reader.
$ \ \ \ \ \ \Box$.  

\vskip 0.2in
{\bf Remark}:  For the sake of completeness, we note that the function
$\sqrt{-q(z^{2})}$ is real and negative for $z \in ( 
\gamma_{1}(a), \gamma_{2}(a))$, and positive imaginary for $z >
\gamma_{2}$.  

\bigskip

Using Lemma \ref{lem:RMTaux}, it follows that as $z$ traverses the
real axis, from $\gamma_{2}$ to $+\infty$, the quantity $-R(z) +
3^{-5/2} \sqrt{-q(z^{2})}$ starts out on the negative real axis, enters the
second quadrant, exiting the second quadrant by crossing the positive
imaginary axis, and then diverges to $\infty$ in the first
quadrant.

In the text following (\ref{eq:51}), we have shown that for $z\in
\mathbb{R}$ and  near $\gamma_{2}$, the appropriate choice of cube
root yielding a function that is analytic is determined by $\phi = 3
\pi$.  As $z$ traverses the real axis from $\gamma_{2}$ to $+\infty$,
this angle must converge to $2 \pi$, and hence we have determined the
behavior of the root $r_{1}(z)$ to be as follows:

\begin{eqnarray}
& & r_{1}(z) = \frac{1}{3} z^{3} +  \frac{1}{3} z^{3} e^{2 i \pi / 3}
\left( 
1 + \frac{i a \sqrt{3}}{z^{3}} + 
- \frac{3}{2 z^{4}} + 
\mathcal{O} \left( \frac{1}{z^{5}}\right)
\right) + \\
\nonumber
& & \hspace{0.8in}+
\frac{1}{3} z^{3} e^{-2 i \pi /3} \left(  
1 - \frac{i a \sqrt{3}}{z^{3}} - \frac{3}{2 z^{4}} + \mathcal{O}
\left( \frac{1}{z^{5}}\right)
\right) \\
\nonumber
& & = - a + \frac{1/2}{z} + \mathcal{O} \left( \frac{1}{z^{2}}\right),
\end{eqnarray}
which establishes (\ref{RMT:root1}).  

We will leave to the motivated reader the proof that the root
$r_{2}$, which is analytic in $\mathbb{C} \setminus [\gamma_{1},
\gamma_{2}]$, possesses the asymptotic expansion (\ref{RMT:root2}) as
$z \to \infty$.

This completes the proof of (3) and (4) of Theorem \ref{thm:3}.

\medskip

{\bf Proof of (5) of Theorem \ref{thm:3}}:  To see that the boundary
relation (\ref{eqn:BVRel1}) holds true, one 
first observes that for $z \in I_{1}$, both boundary values
$r_{1}^{\pm}$ are roots of the cubic (\ref{quart:01}).  Since they are
not both the same root, and since $r_{2}$ is analytic in a
neighborhood of $I_{1}$, it follows that there must be a relation
between $r_{1}^{\pm}$ and $r_{3}^{\pm}$.  The reader may verify that
this monodromy relation is
\begin{eqnarray*}
r_{1}^{\pm} = r_{3}^{\mp} \ \mbox{ for } z \in I_{1}.
\end{eqnarray*}
It is then straightforward to verify that
\begin{eqnarray*}
\left( f_{1}^{-} + f_{1}^{+} \right) + f_{2}  = r_{1}^{+} + r_{1}^{-}
+ r_{2} + a = r_{1}^{+} + r_{2} + r_{3}^{+} + a = z^{3} + a,
\end{eqnarray*}
since the sum of the roots of the cubic (\ref{quart:01}) must be
$z^{3}$.  The boundary relation (\ref{eqn:BVRel2}) follows by very
similar reasoning, the difference being that on $I_{2}$, the boundary
values $r_{2}^{\pm}$  are related to the boundary values of $r_{3}$ as
follows: $r_{2}^{\pm} = r_{3}^{\mp}$.  

{\bf Proof of (6):}  The inequalities (\ref{eqn:MeasReal}) may be seen to be true by one of
the following two approaches.  We will consider $j=1$, as the
considerations for $j=2$ are entirely similar.

\begin{enumerate}
\item Using the expansions (\ref{eq:49}) and (\ref{eq:49b}) (with
  suitable choice of the angle $\phi$), valid
  over the entire interval $I_{1}$, one finds that the difference
  between the boundary values from the upper and lower half planes is
  positive imaginary.

\item One shows that the two roots $r_{1}^{+}$ and $r_{1}^{-}$ are complex
  conjugates of each other, and that $r_{1}^{+}$ has positive
  imaginary part.
\end{enumerate}

{\bf Proof of (7)}:  We now prove that the inequality
(\ref{eqn:VariationIneq}) is true. 
First, recall that on $I_{1}$, (\ref{eqn:BVRel1}) holds true.
Next observe that for $s \in \mathbb{R} \setminus I_{1}$,
\begin{eqnarray}
\label{eq:57}
 f_{1}^{-}(s) + f_{1}^{+}(s) + f_{2}^{-}(s)
- V_{1}'(s)  =  2 f_{1}(s) + f_{2}^{-}(s) - s^{3} - a = r_{1}(s) - r_{3}^{-}(s).
\end{eqnarray}
It turns out that one must verify the inequality
(\ref{eqn:VariationIneq}) separately on each of the intervals 
$(-\infty, -\gamma_{2})$,
$(-\gamma_{1}, \gamma_{1})$, $I_{2}$, and
$(\gamma_{2}, \infty)$.  

Local analysis near the branch points
$-\gamma_{2}$, $-\gamma_{1}$, and $\gamma_{2}$ show that on each of
the intervals $(-\infty, -\gamma_{2})$,
$(-\gamma_{1}, \gamma_{1})$ and $(\gamma_{2}, \infty)$ there are three
real roots.  Since the only places where any two roots can
coincide are the 
branch points $\pm \gamma_{1}$ and $\pm \gamma_{2}$, the quantity
(\ref{eq:57}) is of one sign for $s$ in each of $(-\infty, -\gamma_{2})$,
$(-\gamma_{1}, \gamma_{1})$ and $(\gamma_{2}, \infty)$.  The signature
in each of these intervals can be determined by a (by now)
straightforward local analysis in a vicinity of each branch point.

For example, for $s=-\gamma_{1}$, the two roots
$r_{1}$ and $r_{3}$ coincide, and the following local expansions hold
true:

\begin{eqnarray}
  \label{eq:52}
&& 
\ \ \ \ r_{1}(s) - \frac{1}{3} s^{3} = \\
&&\nonumber \hspace{.5in}
\sqrt[3]{\left| \frac{-R}{2} \right|}  \left[ \cos{ (2 \pi/3)}\left( 1 + \mathcal{O}\left(
    \left| q(s^{2})\right|\right) \right)   \ - \ 
\frac{2 i}{3}  \left(
  \frac{3^{-5/2} \sqrt{-q(s^{2})}}{R(s)} \right) \sin{ (2 \pi/3)}
\left(1 +  \mathcal{O}\left(
    \left| q(s^{2})\right|\right) \right) \right]
,\\
&& 
\ \ \ \ r_{3}(s) - \frac{1}{3} s^{3} = \\
&&\nonumber \hspace{.5in}
\sqrt[3]{\left| \frac{-R}{2} \right|} \left[  \cos{ (4 \pi/3)}\left( 1 + \mathcal{O}\left(
    \left| q(s^{2})\right|\right) \right)   \ - \ 
\frac{2 i}{3}  \left(
  \frac{3^{-5/2} \sqrt{-q(s^{2})}}{R(s)} \right) \sin{ (4 \pi/3)}
\left(1 +  \mathcal{O}\left(
    \left| q(s^{2})\right|\right) \right) \right]
.\end{eqnarray}
Now for $s$ real, $s>-\gamma_{1}$, and $s$ near $-\gamma_{1}$,
$\sqrt{-q(s^{2})} $ is a number on the negative imaginary axis.
Therefore
\begin{eqnarray}
&& 
\ \ \ \ r_{1}(s) - r_{3}(s)  = 
 \ - \ 
\sqrt[3]{\left| \frac{-R}{2} \right|}
\frac{4 i}{3}  \left(
  \frac{3^{-5/2} \sqrt{-q(s^{2})}}{R(s)} \right) \sin{ (2 \pi/3)}
+ \mathcal{O}\left(
    \left| q(s^{2})\right|\right) < 0,
\end{eqnarray}
and then it follows that the inequality (\ref{eqn:VariationIneq}) is
true for $s \in (-\gamma_{1}, \gamma_{1})$.
Similar analysis shows that the same inequality is true on $(-\infty,
-\gamma_{2})$, and we will leave
the details of those calculations to the reader.  

For $z$ in the interval $I_{2}$,  the root $r_{1}$ remains real, but
the root $r_{3}$ (either the ``$+$'' or the ``$-$'' boundary value) is
not real, and the quantity 
$\mbox{Re } \left(r_{1}(z) - r_{3}^{(-)}(z) \right)$ may change signs
without two roots coinciding. 

However, recalling the form (\ref{eq:47}) of the roots of the cubic
(\ref{quart:01}), along with the useful relationship (\ref{eq:50}), we
may represent the roots $r_{j}$ as follows:

\begin{eqnarray}
\label{eq:58}
r_{j} = \frac{1}{3} z^{3} + \ \left( \frac{- R(z) + 3^{-5/2}\sqrt{-q(z^{2})}}{2}
\right)^{1/3} +  
e^{-2 i \phi_{j}/3}
\left(
\frac{-R(z) - 3^{-5/2} \sqrt{-q(z^{2})}}{2}
\right)^{1/3}\ ,
\end{eqnarray}
where the choice of cube-root, and the phase $\phi_{j}$, depends upon
which root is being considered.  

We will need the following properties:
\begin{itemize}
\item  The function
$\sqrt{-q(z^{2})}$ is real and negative for $z \in ( 
\gamma_{1}(a), \gamma_{2}(a))$.

\item The function $R(z)$ is strictly positive for 
$z\in ( \gamma_{1}(a), \gamma_{2}(a))$.

\item The quantities $- R(z) + 3^{-5/2}\sqrt{-q(z^{2})}$ and 
$- R(z) - 3^{-5/2}\sqrt{-q(z^{2})}$ are nonzero for  $z \in [
\gamma_{1}(a), \gamma_{2}(a)]$, and consequently they are both
strictly negative for all $z \in [ \gamma_{1}(a), \gamma_{2}(a)]$.
\end{itemize}

With the above properties, and the choice $\phi_{1} = 3 \pi$ (see the
text following (\ref{eq:51})), we have
\begin{eqnarray}
& & \left( r_{1} - r_{3}^{-}  \right)= 
e^{ i \pi} \left| \frac{- R(z) + 3^{-5/2}\sqrt{-q(z^{2})}}{2}
\right|^{1/3}  +  
 e^{ i \pi}
\left|
\frac{-R(z) - 3^{-5/2} \sqrt{-q(z^{2})}}{2}
\right|^{1/3}  + \\
\nonumber
& &  \ \ \ \ \ \ \ \  -  \left(
e^{ i \phi_{3}/3}\left| \frac{- R(z) + 3^{-5/2}\sqrt{-q(z^{2})}}{2}
\right|^{1/3} +  
e^{-2 i \phi_{j}/3} e^{ i \phi_{j}/3}
\left|
\frac{-R(z) - 3^{-5/2} \sqrt{-q(z^{2})}}{2}
\right|^{1/3}  \right)  \\
\nonumber
& & =  \left( e^{ i \pi} - e^{ i \phi_{3}/3} \right) 
\left| \frac{- R(z) + 3^{-5/2}\sqrt{-q(z^{2})}}{2}
\right|^{1/3} +  \left(  e^{ i \pi}  - e^{- i \phi_{3}/3}\right)
\left|
\frac{-R(z) - 3^{-5/2} \sqrt{-q(z^{2})}}{2}
\right|^{1/3},
\end{eqnarray}
and therefore we have shown that $\mbox{Re } \left( r_{1} - r_{3}^{-}
\right) < 0 $ for all $z \in [\gamma_{1}, \gamma_{2}]$, which
completes the proof that (\ref{eqn:VariationIneq}) is true for $z \in
\gamma_{1}, \gamma_{2})$.  

Lastly, since $\mbox{Re } \left( r_{1} - r_{3}^{-} \right) < 0 $ for 
$z \in [\gamma_{1}, \gamma_{2}]$, and since all three roots are
distinct for all $z \in (\gamma_{2},\infty)$, it follows that this inequality
remains true in $(\gamma_{2},\infty)$ as well, and hence we have
completed the proof that (\ref{eqn:VariationIneq}) holds true for all
$z \in \mathbb{R} \setminus I_{1}$.

The proof that (\ref{eqn:VariationIneq2}) holds true follows by very similar
arguments, and we will omit these details here.

Having completed the proof of (7) of Theorem \ref{thm:3}, we have
completed the proof of the entire Theorem.  $\Box$

{\bf Remark:}  We note that at each of the four endpoints, the roots
$r_{1}$ and $r_{2}$ possess square root singularities.  This implies that
in addition to verifying all of the equalities and inequalities of the
ideal situation A-D, conditions E1 and E2 also hold true.

\section{Complete analysis of the Riemann--Hilbert problem
  \ref{rhp:04} in the ``one-cut'' case}
\label{sect:rhpanalysis}

The starting point of this section is the Riemann--Hilbert problem
(\ref{RHgen2}).  In this section we will show that {\it if} the ideal
situation described in section \ref{AlgCurve} is achieved
(including the ``Airy conditions'' E1 and E2), and if each
of the sets $I_{j}, j=1,2$ is a single interval, then one may obtain a
complete asymptotic description of the solution of the
Riemann--Hilbert problem (\ref{RHgen2}).  Bleher and Kuijlaars
\cite{BK2} have carried out a Riemann--Hilbert analysis for the
Gaussian case, and their work may be applied directly to the present
more general situation.  Therefore, we will only present a summary of the
relevant transformations.  Of course, it should be possible to
generalize the work of this section to the case of arbitrary (finite)
numbers of subintervals comprising the sets $I_{1}$ and $I_{2}$, but
this is putting the cart before the horse as there is at present no existence
theorem in the multi-cut case.  In what follows we will let the
endpoints of the intervals be defined as follows:
\begin{eqnarray}
I_{1} = (\gamma_{1}, \delta_{1}), \ \ \ \ I_{2} = (\gamma_{2}, \delta_{2}),
\end{eqnarray}
and we will use the following particular choices of antiderivatives of
$f_{1}$ and $f_{2}$ to define $g_{1}$ and $g_{2}$:
\begin{eqnarray}
\label{eq:gdef1}
& & g_{1} = \frac{n}{n_{1}}\int_{\delta_{1}}^{z} f_{1}(s) ds - c_{1}\ ,
\\
\label{eq:gdef2}
& & g_{2} = \frac{n}{n_{2}}\int_{\delta_{2}}^{z} f_{2}(s) ds - c_{2}\ ,
\end{eqnarray}
where in (\ref{eq:gdef1}) the contour is chosen in the cut plane
$\mathbb{C} \setminus (-\infty, \delta_{1})$, and in (\ref{eq:gdef2})
the contour is taken in $\mathbb{C} \setminus (-\infty, \delta_{2})$,
and the constants 
$c_{1}$ and $c_{2}$ are chosen so that (\ref{eq:5}) holds true,
namely
\begin{eqnarray}
& & c_{1} = \frac{1}{2} \log{\delta_{1}} \ + \int_{\delta_{1}}^{+\infty}
\left( f_{1}(s) - \frac{1}{2 s}\right) ds, \\
& & c_{2} = \frac{1}{2} \log{\delta_{2}} \ + \int_{\delta_{2}}^{+
  \infty} \left( f_{2}(s) + \frac{1}{2 s} \right) ds.
\end{eqnarray}
(Of course, then $g_{0}$ is determined via (\ref{eq:5a}).)

{\bf Remark:}  In what follows, we will assume a certain type of {\it
  genericity} in the following sense.  We are considering a sequence
of triples, $(n_{1}, n_{2}, n)$ so that $n =n_{1}+n_{2}$ and $n_{j}/n
\rightarrow x_{j}$.  The implicit characterization of the functions
$g_{1}$ and $g_{2}$ in terms of the equation $E(w,z) = 0$ with
$E(w,z)$ defined in (\ref{eq:20}) depends on $x_{1}$ and $x_{2}$, and
we will assume that the ``ideal situation'' holds true not just for
$x_{1}$ and $x_{2}$, but for all values of $(x_{1}',x_{2}')$ near $(x_{1},x_{2})$.

The starting point is to realize that, because all of the conditions
of the ``ideal situation'' are satisfied, the jump matrix appearing in
(\ref{RHgen2}), namely
\begin{eqnarray}
V_{B} =     \begin{pmatrix}
    e^{-n(g_{0}^{+}-g_{0}^{-})} & e^{n g_{0}^{-} +
    n_{1} g_{1}^{+}- n V_{1}-n_{1}\ell_{1}} & 
e^{n g_{0}^{-} + n_{2} g_{2}^{+}- n V_{2} - n_{2}\ell_{2}} \\
    0 & e^{n_{1}(g_{1}^{+}-g_{1}^{-})}      &  0 \\
    0     & 0     & e^{n_{2}(g_{2}^{+}-g_{2}^{-})}
    \end{pmatrix},
\end{eqnarray}
takes on one of four forms, as follows.
\begin{eqnarray}
& & V_{B}(x) = \begin{pmatrix}
1& e^{- nP_{1}(x)} & 
e^{- n P_{2}(x)} \\
    0 & 1 &  0 \\
    0     & 0     & 1
    \end{pmatrix},  \ \ \mbox{for } x \in (-\infty, \gamma_{1})\cup (\delta_{1},\gamma_{2})\cup(\delta_{2}, \infty), \\
& & 
V_{B}(x) = \begin{pmatrix}
    e^{- i n_{2}\theta_{2}(x) } & e^{-n P_{1}(x)} & 
1\\
    0 & 1      &  0 \\
    0     & 0     & e^{i n_{2} \theta_{2}(x) }
    \end{pmatrix}, \ \ \mbox{for } x \in I_{2} \\
& & 
V_{B}(x) = \begin{pmatrix}
    e^{- i n_{1}\theta_{1}(x)} & 1 & 
e^{-n P_{2}(x)} \\
    0 & e^{i n_{1}\theta_{1}(x)}      &  0 \\
    0     & 0     & 1
    \end{pmatrix}, \ \ \mbox{for } x \in I_{1}, 
\end{eqnarray}
where \begin{itemize}
\item $P_{1} =  x_{1} (g_{1}^{+}+g_{1}^{-}) + x_{2} g_{2}-  V_{1}-x_{1}\ell_{1}$
    is strictly negative on 
    $(-\infty,\gamma_{1}) \cup (\delta_{1}, \infty) $, 
\item $P_{2} =x_{2} (g_{2}^{+} + g_{2}^{-}) + x_{1} g_{1} - V_{2} - x_{2}
    \ell_{2} $ is strictly negative on $(-\infty,\gamma_{2}) 
    \cup (\delta_{2}, \infty)$,
\item $\theta_{2}(x) = - i ( g_{2}^{+}(x) - g_{2}^{-}(x)) $ is
  strictly positive and strictly decreasing on $I_{2}$, and possesses
  an analytic continuation (also called $\theta_{2}$) off the
  interval $I_{2}$, 
\item $\theta_{1}(x) = - i ( g_{1}^{+}(x) - g_{1}^{-}(x)) $ is
  strictly positive and strictly decreasing on $I_{1}$, and possesses
  an analytic continuation (also called $\theta_{1}$) off the
  interval $I_{1}$, 
\end{itemize}
\subsection{Second transformation: Opening lenses}
Figure 1 below shows a decomposition of the plane into five regions, formed by
defining four lens shaped regions surrounding the intervals $I_{1}$
and $I_{2}$.  We will define the matrix $D$ in each of these regions,
as shown in Figure 1.  (In the region exterior to the four lens shaped
regions, $D = B$.)  The matrix $D$ is clearly piecewise analytic.
Because it is related directly to the matrix $B$, it also solves a
Riemann--Hilbert problem.  However, we will not write this problem
down, as it is somewhat auxiliary to our final goal.
This definition is quite similar to the definitions (5.2), (5.4) and
(5.6) of \cite{BK2}. 
\medskip

\font\thinlinefont=cmr5
\begingroup\makeatletter\ifx\SetFigFont\undefined%
\gdef\SetFigFont#1#2#3#4#5{%
  \reset@font\fontsize{#1}{#2pt}%
  \fontfamily{#3}\fontseries{#4}\fontshape{#5}%
  \selectfont}%
\fi\endgroup%
\mbox{\beginpicture
\setcoordinatesystem units <.700000cm,.700000cm>
\unitlength=1.00000cm
\linethickness=1pt
\setplotsymbol ({\makebox(0,0)[l]{\tencirc\symbol{'160}}})
\setshadesymbol ({\thinlinefont .})
\setlinear
%
%
\put{$\delta_{2}$} [lB] at 20.637 16.510
%
%
\linethickness= 0.500pt
\setplotsymbol ({\thinlinefont .})
{\color[rgb]{0,0,0}\circulararc 106.260 degrees from 20.320 16.510 center at 17.780 14.605
}%
%
%
\linethickness= 0.500pt
\setplotsymbol ({\thinlinefont .})
{\color[rgb]{0,0,0}\circulararc 106.260 degrees from 15.240 16.510 center at 17.780 18.415
}%
%
%
\linethickness= 0.500pt
\setplotsymbol ({\thinlinefont .})
{\color[rgb]{0,0,0}\circulararc 106.260 degrees from  2.540 16.510 center at  5.080 18.415
}%
%
%
\linethickness= 0.500pt
\setplotsymbol ({\thinlinefont .})
{\color[rgb]{0,0,0}\putrule from  2.540 16.510 to  7.620 16.510
}%
%
%
\linethickness= 0.500pt
\setplotsymbol ({\thinlinefont .})
{\color[rgb]{0,0,0}\putrule from 15.240 16.510 to 20.320 16.510
}%
%
%
\linethickness= 0.500pt
\setplotsymbol ({\thinlinefont .})
{\color[rgb]{0,0,0}\plot  3.175 19.526  4.763 17.145 /
%
%
\plot  4.569 17.321  4.763 17.145  4.674 17.392 /
}%
%
%
\linethickness= 0.500pt
\setplotsymbol ({\thinlinefont .})
{\color[rgb]{0,0,0}\plot  2.857 13.652  5.080 15.875 /
%
%
\plot  4.945 15.650  5.080 15.875  4.855 15.740 /
}%
%
%
\linethickness= 0.500pt
\setplotsymbol ({\thinlinefont .})
{\color[rgb]{0,0,0}\plot 14.446 19.367 17.462 17.145 /
%
%
\plot 17.220 17.245 17.462 17.145 17.296 17.347 /
}%
%
%
\linethickness= 0.500pt
\setplotsymbol ({\thinlinefont .})
{\color[rgb]{0,0,0}\plot 14.129 13.494 17.304 15.875 /
%
%
\plot 17.139 15.672 17.304 15.875 17.062 15.773 /
}%
%
%
\put{$D=B\begin{pmatrix}
1 & 0 & 0\\
-e^{-n(g_{1}^{+}-g_{1}^{-})} & 
 1      & - e^{-n P_{2}} \\
     0& 0     &  1
    \end{pmatrix}
$} [lB] at  2.540 19.685
%
%
\put{$D=B
\begin{pmatrix}
1 & 0 & 
0\\
    0 & 1      &  0 \\
      -e^{-n_{2}(g_{2}^{+}-g_{2}^{-})}0     & -e^{-n P_{1}(x)}     & 1
    \end{pmatrix}$} [lB] at 13.652 19.685
%
%
\put{$D=B\begin{pmatrix}
1& 0 & 
0 \\
e^{n_{1}(g_{1}^{+}-g_{1}^{-})} & 1      &  - e^{- n P_{2} } e^{n_{1}(g_{1}^{+}-g_{1}^{-})} \\
    0     & 0     & 1
    \end{pmatrix},$} [lB] at  2.381 13.018
%
%
\put{$D=B\begin{pmatrix}
1 & 0 & 
0\\
    0 & 1      &  0 \\
e^{n_{2}(g_{2}^{+}-g_{2}^{-})}       & - e^{-n P_{1}(x)}   e^{n_{2}(g_{2}^{+}-g_{2}^{-})}  & 1
    \end{pmatrix}$} [lB] at 13.494 13.018
%
%
\put{$\delta_{1}$} [lB] at  7.938 16.351
%
%
\put{$\gamma_{1}$} [lB] at  1.746 16.351
%
%
\put{$\gamma_{2}$} [lB] at 14.722 16.510
%
%
\linethickness= 0.500pt
\setplotsymbol ({\thinlinefont .})
{\color[rgb]{0,0,0}\circulararc 106.260 degrees from  7.620 16.510 center at  5.080 14.605
}%
\linethickness=0pt
\putrectangle corners at  1.746 20.015 and 20.637 12.895
\endpicture}

\smallskip
\begin{itemize}
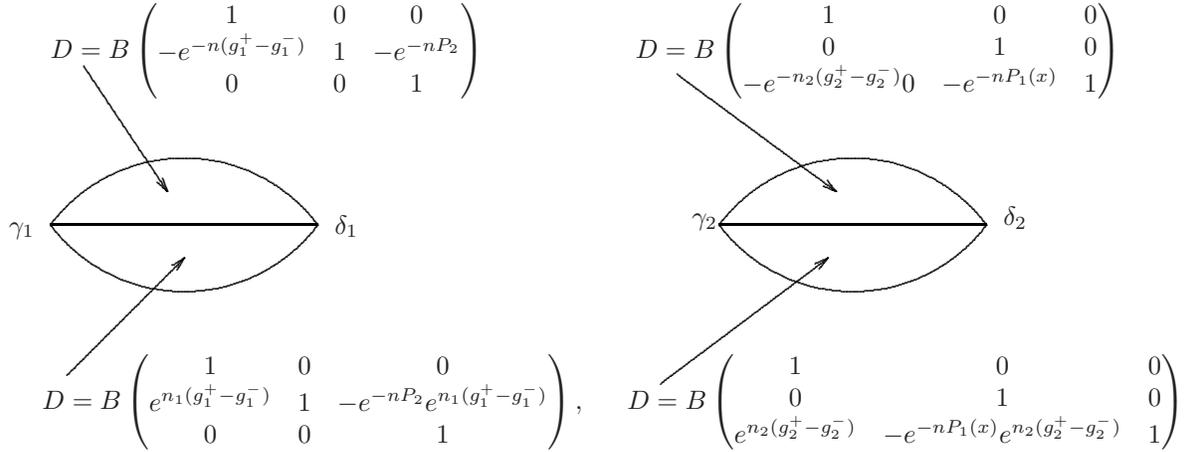
\item[]{\bf Figure 1.}  The decomposition of the plane into 5
  regions, the interior of the 4 lens shaped regions surrounding the
  intervals $(\gamma_{1}, \delta_{1})$ and $(\gamma_{2},
  \delta_{2})$, and the one exterior region.  The matrix $D$ is
  defined as shown in each of the 4 bounded regions, and $D=B$ for $z$
  exterior to all four lens shaped regions.
\end{itemize}

{\bf Remark:} In order to define the matrix $D$, one must extend, for
$j=1,2$, the quantities $g_{j}^{+} - g_{j}^{-}$ and $P_{j}$, for $j=1,2$ off
the interval $I_{j}$.  The justification for this uses the connection
to the algebraic curve.  For example, 
\begin{eqnarray}
g_{2}^{+}(z) - g_{2}^{-}(z) = \frac{1}{x_{2}}
\int_{\delta_{2}}^{z}f_{2}^{+}-f_{2}^{-} dx = \int_{\delta_{2}}^{z}f_{2}^{+}-f_{0}^{+} dx = \int_{\delta_{2}}^{z}f_{0}^{-}-f_{2}^{-} dx,
\end{eqnarray}
which clearly demonstrates the local analyticity of
$g_{2}^{+}-g_{2}^{-}$.  The other quantities are seen to be analytic
by similar calculations.
\medskip

The next step is to define a piecewise analytic matrix valued function
$D_{approx}(z)$ which, it will be shown, is a globally uniform
approximation to the matrix valued function $D$.  We will require the
following auxiliary functions.  Set 
\begin{eqnarray}
&&P_{\gamma_{1}} = r_{1}(\gamma_{1}), \ \ \ \ P_{\delta_{1}} =
r_{1}(\delta_{1}), \\
&&P_{\gamma_{2}} = r_{2}(\gamma_{2}), \ \ \ \ P_{\delta_{2}} =
r_{2}(\delta_{2}). \\
\end{eqnarray}
Let $\Gamma_{1}^{+}$ denote the image of the $+$-side of the
interval $I_{1}$ under the transformation $\xi(z) = r_{0}(z)$ (recall
that $r_{0}(z)$ has branch cuts along $I_{1}$ and $I_{2}$), and let
$\Gamma_{2}$ denote the image of the $+$-side of the interval $I_{2}$
under the same transformation $\xi(z)=r_{0}(z)$. The reader may verify
that for each $j=1,2$, $\Gamma_{j}^{+}$ is an arc in the upper half
plane connecting $P_{\gamma_{j}}$ and $P_{\delta_{j}}$.

Now we define three functions, $M_{j}(z)$ ($j=1,2,3$) as follows
(cf. (6.9) of \cite{BK2}).
\begin{eqnarray}
&&M_{1}(\xi) = \frac{ \xi^{2} - a^2}{\sqrt{\left(\xi - P_{\gamma_{1}}\right)
\left(\xi - P_{\delta_{1}}\right)\left(\xi-P_{\gamma_{2}}\right)\left(\xi -
    P_{\delta_{2}}\right)}}, \\
&&M_{2}(\xi) = c_{2}\frac{ \xi + a}{\sqrt{\left(\xi - P_{\gamma_{1}}\right)
\left(\xi - P_{\delta_{1}}\right)\left(\xi-P_{\gamma_{2}}\right)\left(\xi -
    P_{\delta_{2}}\right)}}, \\
&&M_{3}(\xi) = c_{3}\frac{ \xi - a}{\sqrt{\left(\xi - P_{\gamma_{1}}\right)
\left(\xi - P_{\delta_{1}}\right)\left(\xi-P_{\gamma_{2}}\right)\left(\xi -
    P_{\delta_{2}}\right)}}, 
\end{eqnarray}
where in each instance, the quantity $\sqrt{\left(\xi - P_{\gamma_{1}}\right)
\left(\xi - P_{\delta_{1}}\right)\left(\xi-P_{\gamma_{2}}\right)\left(\xi -
    P_{\delta_{2}}\right)}$ is taken to be analytic in $\mathbb{C}
\setminus \left(\Gamma_{1}^{+} \cup \Gamma_{2}^{+} \right)$, and the
constants $c_{1}$ and $c_{2}$ are taken to be 
\begin{eqnarray}
&&c_{1} = \frac{1}{2 a} \sqrt{\left(a - P_{\gamma_{1}}\right)
\left(a - P_{\delta_{1}}\right)\left(a-P_{\gamma_{2}}\right)\left(a -
    P_{\delta_{2}}\right)}, \\
&&c_{2} = \frac{-1}{2a} \sqrt{\left(-a - P_{\gamma_{1}}\right)
\left(-a - P_{\delta_{1}}\right)\left(-a-P_{\gamma_{2}}\right)\left(-a -
    P_{\delta_{2}}\right)}.
\end{eqnarray}

We will require the following $3 \times 3$ matrix $P(z)$:
\begin{eqnarray}
& &P(z) = \begin{pmatrix}
M_{1}(r_{0}(z)) & M_{1}(r_{1}(z)) & M_{1}(r_{2}(z)) \\
M_{3}(r_{0}(z)) & M_{3}(r_{1}(z)) & M_{3}(r_{2}(z)) \\
M_{2}(r_{0}(z)) & M_{2}(r_{1}(z)) & M_{2}(r_{2}(z))
\end{pmatrix}.
\end{eqnarray}
For future reference, the matrix $P(z)$ solves the following
Riemann--Hilbert problem.

\begin{rhp}

\label{rhp:model}

\smallskip

\begin{enumerate}
\item[(a)] $P$ is analytic on $\mathbb{C} \setminus \left( I_{1} \cup I_{2}\right)$. 
\item[(b)] The boundary values of $P$ satisfy 
\begin{eqnarray} 
&&    P_+(x) = P_-(x)
    \begin{pmatrix}
0&1&0\\
-1&0&0\\
0&0&1
    \end{pmatrix}, \ \mbox{for } z \in I_{1}, \\
&&
P_+(x) = P_-(x)
    \begin{pmatrix}
0&0&1\\
0&1&0\\
-1&0&0
    \end{pmatrix}, \ \mbox{for } z \in I_{2}, \\
\end{eqnarray}
\item[(c)] As $z \to \infty$, we have
\begin{equation} 
    P(z) = \left(I + O\left(\frac{1}{z}\right)\right).
\end{equation}
\end{enumerate}

\end{rhp}

\medskip

Next we draw 4 small circles, centered at each of the endpoints
$\gamma_{1}, \delta_{1}, \gamma_{2}, \delta_{2}$, and decompose the
plane again, as shown in Figure 2.  
\medskip

\font\thinlinefont=cmr5
\begingroup\makeatletter\ifx\SetFigFont\undefined%
\gdef\SetFigFont#1#2#3#4#5{%
  \reset@font\fontsize{#1}{#2pt}%
  \fontfamily{#3}\fontseries{#4}\fontshape{#5}%
  \selectfont}%
\fi\endgroup%
\mbox{\beginpicture
\setcoordinatesystem units <.700000cm,.700000cm>
\unitlength=1.00000cm
\linethickness=1pt
\setplotsymbol ({\makebox(0,0)[l]{\tencirc\symbol{'160}}})
\setshadesymbol ({\thinlinefont .})
\setlinear
%
%
\put{$\gamma_{1}$%
} [lB] at  2.223 16.034
%
%
\linethickness= 0.500pt
\setplotsymbol ({\thinlinefont .})
{\color[rgb]{0,0,0}\ellipticalarc axes ratio  0.794:0.794  360 degrees 
	from  8.414 16.510 center at  7.620 16.510
}%
%
%
\linethickness= 0.500pt
\setplotsymbol ({\thinlinefont .})
{\color[rgb]{0,0,0}\ellipticalarc axes ratio  0.794:0.794  360 degrees 
	from 16.034 16.510 center at 15.240 16.510
}%
%
%
\linethickness= 0.500pt
\setplotsymbol ({\thinlinefont .})
{\color[rgb]{0,0,0}\ellipticalarc axes ratio  0.794:0.794  360 degrees 
	from 21.114 16.510 center at 20.320 16.510
}%
%
%
\linethickness= 0.500pt
\setplotsymbol ({\thinlinefont .})
{\color[rgb]{0,0,0}\putrule from  2.540 16.510 to  7.620 16.510
}%
%
%
\linethickness= 0.500pt
\setplotsymbol ({\thinlinefont .})
{\color[rgb]{0,0,0}\putrule from 15.240 16.510 to 20.320 16.510
}%
%
%
\put{$\delta_{1}$%
} [lB] at  7.461 16.034
%
%
\put{$\gamma_{2}$%
} [lB] at 14.764 16.034
%
%
\put{$\delta_{2}$%
} [lB] at 20.161 16.034
%
%
\linethickness= 0.500pt
\setplotsymbol ({\thinlinefont .})
{\color[rgb]{0,0,0}\ellipticalarc axes ratio  0.794:0.794  360 degrees 
	from  3.334 16.510 center at  2.540 16.510
}%
\linethickness=0pt
\putrectangle corners at  1.729 17.321 and 21.131 15.699
\endpicture}

\medskip
\begin{itemize}
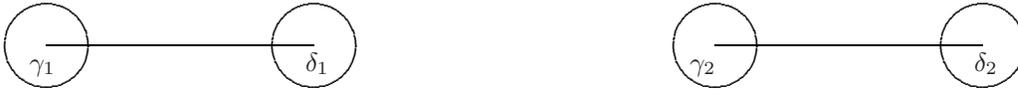
\item[]{\bf Figure 2.}  Four disks, centered at each of the endpoints
  of the intervals $I_{1}$ and $I_{2}$.  We decompose the plane into
  five regions:  the four disks and the one exterior region.
\end{itemize}
\medskip

We will now define a matrix $D_{approx}$ which will be a uniformly
valid approximation to the matrix $D$.  The matrix $D_{approx}$ will
be defined separately in each of the 5 regions as shown in Figure 2.
We set
\begin{eqnarray}
D_{approx}(z) = P(z), \ \ \mbox{ for } z \mbox{ outside the four disks}.
\end{eqnarray}

For $z$ in the interior of each of the four disks, $D_{approx}$ is
defined in a (by now) standard way using the solutions of the Airy
equation.  Rather than presenting the details for each of the disks,
we will focus only on the disk centered at $\delta_{1}$, as the
construction for each of the other disks may be carried out by very
similar calculations.  Define
\begin{eqnarray}
S(z) = \left[  \frac{3}{4} \int_{\delta_{1}}^{z} \left( 2 f_{1}(x) +
    f_{2}(x)  - V_{1}'(x) \right)  dx 
\right]^{2/3},
\end{eqnarray}
taken to be analytic in a neighborhood of $z = \delta_{1}$, and with
$S'(\delta_{1}) > 0$ (which is possible because of conditions E1 and
E2).  It turns out that with this definition, one may 
take an appropriate branch so that $S(z)^{3/2} = -  \frac{3 i }{4}
\theta_{1}(z) $ for $z$ in the upper lens shaped region
above $I_{1}$, and $S(z)^{3/2} =  \frac{3 i }{4}
\theta_{1}(z) $ for $z$ in the lower lens shaped region
below $I_{1}$.

Next, define the matrix $\Phi(S)$ as follows
(cf. \cite[(7.13)]{BK2}).
\begin{eqnarray}
\Phi(s) = \left\{
\begin{array}{cc}
\begin{pmatrix}
y_{0}(s)&-y_{2}(s)&0\\
y_{0}'(s)&-y_{2}'(s)&0\\
0&0&1\\
\end{pmatrix}, & \mbox{ for } 0 < \arg{s} < \frac{2\pi}{3},\\
\begin{pmatrix}
-y_{1}(s)&-y_{2}(s)&0\\
-y_{1}'(s)&-y_{2}'(s)&0\\
0&0&1\\
\end{pmatrix}, & \mbox{ for } \frac{2 \pi}{3} < \arg{s} < \pi,\\
\begin{pmatrix}
-y_{2}(s)&y_{1}(s)&0\\
-y_{2}'(s)&y_{1}'(s)&0\\
0&0&1\\
\end{pmatrix}, & \mbox{ for } -\pi < \arg{s} < -\frac{2 \pi}{3},\\
\begin{pmatrix}
y_{0}(s)&y_{1}(s)&0\\
y_{0}'(s)&y_{1}'(s)&0\\
0&0&1\\
\end{pmatrix}, & \mbox{ for } -\frac{2 \pi}{3} < \arg{s} < 0.
\end{array}
\right. \ ,
\end{eqnarray}
where $y_{0}, y_{1}$, and $y_{2}$ are the following solutions of the
Airy equation (recall that $\mbox{Ai}(x)$ is the unique solution to
Airy's equation $y''=x y$ which is exponentially decaying for $x \to
+\infty$):
\begin{eqnarray}
y_{0}(s) = \mbox{Ai}(s), \ \ \ \ y_{1}(s) = \omega \mbox{Ai}(\omega
s), \ \ \ \ 
y_{2}(s) = \omega^{2} \mbox{Ai}(\omega^{2} s)
\end{eqnarray}
(here $\omega = e^{ \frac{2 \pi i}{3}}$).

And then define $E_{n}(z)$ for $z$ in the disk centered at
$\delta_{1}$ (again following \cite{BK2}) as follows:
\begin{eqnarray}
E_{n}(z) = \sqrt{\pi} P(z) \begin{pmatrix}
1&-1&0\\
-i&-i&0\\
0&0&1\\
\end{pmatrix}
\begin{pmatrix}
n^{1/6}S(z)^{1/4}&0&0\\
0&n^{-1/6}S(z)^{-1/4}&0\\
0&0&1\\
\end{pmatrix}.
\end{eqnarray} 
Finally then, we may define $D_{approx}(z)$  for $z$ in the disk
centered at $\delta_{1}$:
\begin{eqnarray}
D_{approx}(z) = E_{n}(z) \Phi \left( n^{2/3} S(z) \right) \
\mbox{diag} \left( e^{-\frac{n_{1}}{2} \left(g_{1}^{+}-g_{1}^{-}
    \right)}, \ e^{\frac{n_{1}}{2} \left(g_{1}^{+}-g_{1}^{-}
    \right)}, \ 1 \ \right).
\end{eqnarray}
There is a similar definition of $D_{approx}$ in each of the other 3
disks shown in Figure 2, which are entirely similar to the
construction presented here \cite{BK2}.  We will omit these
definitions here.

\medskip

The matrix $D_{approx}$ is a piecewise analytic function in
the entire plane, and it is a global approximation to the matrix
$D(z)$ defined in Figure 1.  The proof of this follows a by now
straightforward procedure, which has been described in
\cite[(8.1)-(8.6)]{BK2}.  Indeed, setting
\begin{eqnarray}
R(z) = D(z) D_{approx}^{-1}(z),
\end{eqnarray}
one verifies that $R(z)$ satisfies a Riemann--Hilbert problem of
exactly the same type as the analogous function $R(z)$ defined in
\cite[(8.1)]{BK2}:  jumps across the boundaries of the disks are $I +
\mathcal{O}(n^{-1})$, and jumps across all other contours are $I +
\mathcal{O}\left( e^{- c n}\right)$, and in addition the actual
contours chosen may be deformed slightly (as in the text between (8.5)
and (8.6) of \cite{BK2}).  Therefore one may conclude that 
\begin{eqnarray}
R(z) = I + \mathcal{O} \left(  \frac{1}{n ( |z| + 1)}\right) \
\mbox{as } n \to \infty,
\end{eqnarray}
uniformly for $z \in \mathbb{C}$.

\section{Asymptotics for eigenvalue statistics of  random matrices
  with source}
\label{results}

In this short section we will describe some of the results that follow
immediately from the results of the previous section.  The proofs of
these results follow by considerations entirely similar to those
carried out in the Gaussian case by Bleher and Kuijlaars \cite{BK2}:
one expresses the kernel $K_{n}(x,y)$ in terms of the explicit
transformations leading to $R(z)$, whose asymptotic expansion for $n
\to \infty$ is under control.

\begin{theorem}\label{thm:rmts01}
Suppose that the conditions described as the ``ideal situation'',
items A-D, described in Section 2 are satisfied, along with the
conditions described in E1 and E2 of Section 3.  Suppose in addition
that the sets $I_{1}$ and $I_{2}$ are each single intervals, with
$I_{1} = (\gamma_{1},\delta_{1})$ and $I_{2}=(\gamma_{2},
\delta_{2})$.  Then the following results hold true.
\begin{itemize}
\item
The mean density of states, $\rho_{n}(x) := K_{n}(x,x)$ converges, as
$n \to \infty$, to a limiting mean density, supported on $I_{1} \cup
I_{2}$, with
\begin{eqnarray}
\lim_{n \to \infty} \rho_{n}(x) = - \frac{1}{2 \pi i} \frac{d}{dx}
\left( g_{j}^{+}(x) - g_{j}^{-}(x) \right), \ \ \mbox{for } x \in
I_{j}, \ \ j = 1,2.
\end{eqnarray}
\item Bulk universality holds true:  for every $x_{0} \in (\gamma_{1},
  \delta_{1}) \cup (\gamma_{2}, 
  \delta_{2})$, and every $u,v \in \mathbb{R}$, we have 
\begin{eqnarray}
\lim_{n \to \infty} \frac{1}{ n \rho(x_{0})} K_{n}\left(x_{0} + \frac{u}{n
  \rho(x_{0})}, x_{0} + \frac{v}{n \rho(x_{0})}\right) = \frac{\sin{
  \pi ( u - v)}}{\pi (u-v)}.
\end{eqnarray}
\item Edge universality also holds true, at each endpoint of $I_{1}
  \cup I_{2}$:  There are constants $c_{\gamma_{1}}, c_{\delta_{1}},
  c_{\gamma_{2}},$ and $c_{\delta_{2}}$ so that for $j=1,2$, the
  following statements hold true:  for every $u,v \in
  \mathbb{R}$, we have
\begin{eqnarray}
& & \lim_{n \to \infty} \frac{1}{ (c_{\gamma_{j}}n)^{2/3}} K_{n}\left(\gamma_{j}
  + 
\frac{u}{(c_{\gamma_{j}} n)^{2/3}}, \gamma_{j}
  + 
\frac{v}{(c_{\gamma_{j}} n)^{2/3}}\right) = \frac{\mbox{Ai}(u) \mbox{Ai}'(v)
- \mbox{Ai}(v) \mbox{Ai}'(u)}{u-v}, \\
& & \lim_{n \to \infty} \frac{1}{ (c_{\delta_{j}}n)^{2/3}} K_{n}\left(\delta_{j}
  - 
\frac{u}{(c_{\delta_{j}}n)^{2/3}}, \delta_{j}
  - 
\frac{v}{(c_{\delta_{j}}n)^{2/3}}\right) = \frac{\mbox{Ai}(u) \mbox{Ai}'(v)
- \mbox{Ai}(v) \mbox{Ai}'(u)}{u-v}, 
\end{eqnarray}
\end{itemize}

\end{theorem}

{\bf Remark:}  As proven in Section \ref{ExistQuad}, all of the conditions A-D
along with E1 and E2 hold true  for all $a$ sufficiently large, 
for the quartic case $V(x) = \frac{1}{4}
x^{4}$, with $n$ even and $n_{1} = n_{2} = n/2$.  Thus bulk and edge
universality have been extended beyond the Gaussian case for random
matrices with source.  

It is natural to expect that the same results
should hold true for all $a$ sufficiently large, under the more
general assumption that the external field $V(x)$ is convex, real
analytic, and with sufficient growth for $|x| \to \infty$.  While it
seems a rather daunting task to carry out the requisite asymptotic
analysis for $a \to \infty$ to prove that a suitable curve exists
(note that it need not be algebraic), there is a possibility to
combine the analysis described herein with an analysis of the coupled
variational problem described following formula (1.15) in \cite{BK2}.

\section*{Acknowledgement} 

The research of K.T.-R.M.\ was supported, in part, by the National
Science Foundation under grants DMS-0451495 and DMS-0200749.  It is a
great pleasure to thank Marco Bertola, Pavel Bleher, Percy Deift, Nick
Ercolani, Bertrand Eynard, Arno Kuijlaars and Carlos Tomei for
interesting and useful discussions.

\end{document}